\documentclass[aps,preprint,preprintnumbers,amsmath,amssymb]{revtex4}
\usepackage{mathrsfs}
\usepackage{amssymb}
\usepackage{graphicx}
\usepackage{dcolumn}
\usepackage{bm}
\usepackage{color}
\usepackage{setspace}
\usepackage[breaklinks=true,colorlinks=true,linkcolor=blue,urlcolor=blue,citecolor=blue]{hyperref}
\UseRawInputEncoding   
\usepackage{soul}
\usepackage{ulem}
\usepackage{ragged2e}
\makeatletter

\newcommand{\Rmnum}[1]{\expandafter\@slowromancap\romannumeral #1@}

\begin{document}

\title{Multiple superconducting phases and order-parameter evolution in pressurized UTe$_2$}

\author{Shuo Zou$^{1*}$, Fengrui Shi$^{2*}$, Zhuolun Qiu$^{1}$\footnote[1]{These authors contribute equally to this work}, Jia-Long Zhang$^{3}$, Yan Zhang$^2$, Weilong Qiu$^{2}$, Zhuo Wang$^{1}$, Hai Zeng$^{1}$, Yinina Ma$^{4}$, Zheyu Wu$^{5}$, Andrej Cabala$^{6}$, Michal Vali\v{s}ka$^{6}$, Ning Li$^{7}$, Zihan Yang$^{2}$, Kaixin Ye$^{2}$, Jiawen Zhang$^{2}$, Yanan Zhang$^{2}$, Kangjian Luo$^{1}$, Binbin Zhang$^{7}$, Alexander G. Eaton$^{5}$, Chaofan Zhang$^{7}$, Gang Li$^{4}$, Jianlin Luo$^{4}$, Wen Huang$^{3}$\footnote[2]{Corresponding author: Mail address: huangwen@gbu.edu.cn}, Huiqiu Yuan$^{2,8,9}$\footnote[3]{Corresponding author: Mail address: hqyuan@zju.edu.cn}, Xin Lu$^{2,8,9}$\footnote[4]{Corresponding author: Mail address: xinluphy@zju.edu.cn}, and Yongkang Luo$^{1,10}$\footnote[5]{Corresponding author: Mail address: mpzslyk@gmail.com}}

\address{\small{$^1$Wuhan National High Magnetic Field Center and School of Physics, Huazhong University of Science and Technology, Wuhan 430074, China;}}
\address{\small{$^2$Center for Correlated Matter and School of Physics, Zhejiang University, Hangzhou 310058, China;}}
\address{\small{$^3$School of Physical Sciences, Great Bay University, Dongguan 523000, China;}}
\address{\small{$^4$Beijing National Laboratory for Condensed Matter Physics, Institute of Physics, Chinese Academy of Sciences, Beijing 100190, China;}}
\address{\small{$^5$Cavendish Laboratory, University of Cambridge, Cambridge CB3 0HE, United Kingdom;}}
\address{\small{$^6$Charles University, Faculty of Mathematics and Physics, Department of Condensed Matter Physics, Prague 2 121 16, Czech Republic;}}
\address{\small{$^7$College of Advanced Interdisciplinary Studies and Nanhu Laser Laboratory, National University of Defense Technology, Changsha, Hunan 410073, China;}}
\address{\small{$^8$Institute of Fundamental and Transdisciplinary Research, Zhejiang University, Hangzhou 310058, China;}}
\address{\small{$^{9}$Collaborative Innovation Center of Advanced Microstructures, Nanjing 210093, China;}}
\address{\small{$^{10}$State Key Laboratory of Advanced Electromagnetic Technology, Huazhong University of Science and Technology, Wuhan 430074, China.} }

\date{\today}

\maketitle

\textbf{
The recently discovered heavy-fermion spin-triplet superconductor candidate UTe$_2$ provides a rich platform for unconventional pairing and topological phenomena. However, limited has been known about its superconducting order parameters and their evolution with control parameters, largely due to the lack of appropriate symmetry-sensitive detections. Here, we report comprehensive point-contact spectroscopy measurements of pressurized UTe$_2$ on the (0~0~1) surface. The observation of Andreev bound states strongly suggests the presence of a $p_z$ component in the superconducting order parameters. Quantitative analysis based on an extended Blonder-Tinkham-Klapwijk model unveils the superconducting order parameters with a finite odd-$k_z$ component (e.g. $B_{2u}$ or $B_{3u}$) for both ambient and pressurized UTe$_2$. Remarkably, the multiple superconducting phases can be distinguished by a single parameter $\langle \Delta_{z}\rangle/\langle\Delta_{x(y)}\rangle$, the relative weight between the $p_z$-wave and $p_{x(y)}$-wave pairings. These findings place stringent constraints on the pairing symmetry and provide essential spectroscopic signatures for distinguishing pressure-induced multiple superconducting phases in UTe$_2$. }\\

\textbf{INTRODUCTION}

The classification of superconductors is fundamentally governed by the pairing symmetry of their order parameters. Unlike spin-singlet superconductivity (SC) (total spin $S=0$, even parity), which includes both conventional $s$-wave and many unconventional (e.g., $d$-wave) cases, spin-triplet pairing ($S=1$, odd parity) ！ where electrons with parallel spins form Cooper pairs ！ represents a major frontier in condensed matter physics. This distinction is profound because pairing symmetry not only dictates the microscopic mechanism of SC, but also constraints on their physical properties and hence application prospects. Crucially, spin-triplet superconductors may host Majorana zero modes (a kind of non-Abelian quasiparticles localized at topological defects or edges) that constitute a critical resource for fault-tolerant topological quantum computation \cite{AliceaJ-Nature2012,BeenakkerCWJ-ARCP2013,SatoM-RPP2017}. A central but often unresolved challenge, however, is definitive identification of the order parameter of spin-triplet superconductors (or candidates), many of which (e.g. Sr$_2$RuO$_4$, UPt$_3$, etc) have been in long-standing controversy \cite{Mackenzie-Sr2RuO4_npjQM2017, APustogow-Sr2RuO4_Nature2019, RJoynt-UPt3_RMP, DAoki-USC_JPSJ2019}. In particular, analogous to the superfluid $^3$He \cite{LeggettAJ-RMP1976,KojimaH-JPSJ2008}, the interplay between spin and orbital degrees of freedom can stabilize multiple superconducting phases under different conditions \cite{AHuxley-UPt3_Nature2000,FLevy-URhGe_Science2005, SKhim-CeRh2As2_Science2021,GQZheng-K2Cr3As3_JPCS2023}. This renders the determination of SC order parameters even more challenging in that many routine experimental probes may lose their efficacy at the extremes of ultra-low temperature, high pressure ($p$), or strong magnetic field ($\mu_0\mathbf{H}$).

UTe$_2$, standing out as a promising candidate of spin-triplet superconductor, has attracted tremendous recent attention \cite{RanS-Science2019,AokiD-JPSJ2019,XuY-PRL2019,DuanC-Nature2021,HayesI-Nature2021,RosuelA-PRX2023,WuZ-PNAS2024,WuZ-PNAS2025,LiZ-PNAS2025}. Albeit with a relatively low critical transition temperature $T_c$ ($\approx 2$ K), it exhibits exotic superconducting properties, including upper critical fields far exceeding the Pauli limit along all principal axes \cite{RanS-Science2019}, halo-structured reentrance of SC under ultra-high magnetic field \cite{RanS-NP2019,WuZ-PRX2025,LewinSK-Science2025}, possible chirality and time-reversal symmetry breaking \cite{JiaoL-Nature2020, Bae-ChiralNC2021, HayesI-Nature2021}, pair density wave \cite{GuQ-Nature2023}, strange metallicity \cite{ThomasSM-SciAdv2020,WeinbergerTI-Arxiv2025}, etc. The pressure-induced phase diagram of UTe$_2$ is intriguing, too. Three SC phases denoted by SC1-3 were reported under pressure \cite{AokiD-JPSJ2020,ThomasSM-SciAdv2020,WuZ-PRL2025}. Although the boundaries among these SC phases seem to be clear, whether these states are mutually independent, what their pairing symmetries are, and how their order parameters evolve with temperature and pressure remain fundamental open questions. Unfortunately, little is known about these issues due to the lack of symmetry-sensitive experiments under pressure to date \cite{KinjoK-SciAdv2023}.

A natural approach to addressing this challenge is to look into the gap structure and the associated in-gap Andreev bound state (ABS) on specific sample surface. As illustrated in Fig.~1, for an ABS to form at a surface with specular reflection condition, there must be a phase change between the incident and reflected quasiparticle wavefunctions to incur a destructive interference. This can be realized if one or all components of the gap function $\Delta(\mathbf{k})$ changes sign upon going from $\mathbf{k}$ to $\mathbf{k}^\prime$. Such an ABS can be detected through point-contact spectroscopy (PCS) ！ a technique well-suited for extreme conditions ！ by measuring the differential conductance ($G_\text{NS}$) with an appropriate geometric alignment between tunneling current $\mathbf{I}$ and inter-nodal scattering wavevector $\mathbf{q}_\text{N}$, seeing Fig.~1(c-d). In this way, one assumes that the four possible spin-triplet pairings in UTe$_2$ (cf. Table 1) can be distinguished, in part, because zero-energy ABS (ZABS) can show up on the (0 0 1) surface only in the $B_{2u}$ and $B_{3u}$ irreducible representations (IRs) whose order parameters contain the $k_z$ component; $A_u$ may also exhibit in-gap ABS on this surface, however, due to the fully-gapped dispersion, these ABSs are off-center.

Herein, by systematic PCS investigations of UTe$_2$ under hydrostatic pressure, signatures of ZABS are observed on the (0~0~1) surface, unambiguously suggesting the presence of $p_z$ component in the gap function. Fitting the results to an extended Blonder-Tinkham-Klapwijk (BTK) model, we further find that the SC pairings of UTe$_2$ most likely fall in the $B_{2u}$ or $B_{3u}$ representation, and the different SC phases under pressure can be distinguished by a single parameter $\frac{\langle \Delta_z \rangle}{\langle \Delta_{x(y)} \rangle}$, the relative weight between the $p_z$-wave and $p_{x(y)}$-wave pairings. Our study therefore offers an important clue to pinpoint the pairing symmetry in this compound, and also provides crucial spectroscopic evidence for the existence of multiple SC phases therein. \\

\textbf{RESULTS}

PCS is a powerful probe for the density of states and hence the gap structure of superconducting pairing \cite{DuifAM-1989,DeutscherG-RMP2005,DagheroD-SST2010}. At the interface between a normal metal (NM) and a superconductor, an electron incident from the NM is either specularly reflected or retro-reflected as a hole by creating a Cooper pair in the superconductor, the latter of which is the well-known Andreev reflection, cf. Supplementary Fig.~5 in \textbf{Supplementary Information} (\textbf{SI}). 
A schematic set-up for the PCS measurements of UTe$_2$ under hydrostatic pressure is illustrated in Fig.~2(a). To avoid Ag-Te reaction, we used Au epoxy instead. The contact points were located on the (0~0~1) surface, with the magnetic field applied along the $\mathbf{a}$-axis. To verify the sample quality, the standard resistivity measurements were also conducted on the same sample, with electrical current along $\mathbf{a}$. More details about experimental method can be found in the \textbf{Methods} section.

Supplementary Figure 1(a) shows that the $\rho(T)$ profile of our sample S1 at ambient pressure is in good agreement with previously reported results for ultra-clean single crystals grown by the molten soft flux method \cite{SakaiH-PRM2022,EatonAG-NC2024}, exhibiting a high residual resistivity ratio (RRR) up to 410 and high superconducting transition temperatures $T_\text{c}^0=2.1$ K and $T_\text{c}^\text{on}=2.6$ K. The low-$T$ part of $\rho(T)$ under various pressures are displayed in Supplementary Fig.~1(b-c), and a phase diagram is constructed by the false-color contour plot of $\rho(p, T)$ in Fig.~2(b). At low pressure in the regime denoted SC1 ($p<0.3$ GPa), $T_c$ is slightly suppressed with increasing pressure. Beyond this threshold, the resistive $T_c$ turns up, and this yields a dome structure in the intermediate pressure regime $0.3<p\leq1.4$ GPa; however, earlier bulk measurements including AC calorimetry and AC susceptibility measurements both revealed a dual-transition regime: a high-temperature phase (SC2) and a low-temperature phase (SC3) \cite{AokiD-JPSJ2020,ThomasSM-SciAdv2020,WuZ-PRL2025}. The boundaries among these SC phases are depicted by the dashed lines. For pressure above 1.4 GPa, two anomalies denoted by $T_\text{m1}$ and $T_\text{m2}$ appear just above $T_c$, characteristic of the magnetically ordered phases MO1 and MO2, respectively [see Supplementary Fig.~1(c)]. $T_\text{m1}$ increases rapidly with pressure, while $T_\text{m2}$ is weakly pressure dependent. Overall, the as-obtained phase diagram resembles those reported in the literature \cite{ThomasSM-SciAdv2020,KinjoK-SciAdv2023,WuZ-PRL2025}.

Turning now to the PCS results of UTe$_2$. We start with the data at ambient pressure, as shown in Fig.~2(c). The normal-state spectrum is nearly flat. At 2.5 K, just below $T_\text{c}^\text{on}$, a slight hump appears in the differential conductance spectrum. As temperature decreases further, the hump stabilizes, forming a zero-bias conductance peak (ZBCP). Before relating such ZBCP to the feature of unconventional SC states \cite{TanakaY-PRL1995}, it is necessary to rule out other extrinsic origins. First, spurious ZBCP can be caused by heating effect in the thermal regime \cite{YGNaidyuk-PCSBook,SheetG-PRB2004,SSasaki-CuxBi2Se3PRL2011}, in the regard that an increasing bias voltage causes the local current to exceed the critical current, and consequently $G_\text{NS}$ decreases by forming a peak centering about $V=0$. In this case, it is speculated that the amplitude of the peak shall be nearly temperature independent, while the width of the peak shall shrink rapidly with magnetic field. Here, we rule out this possibility because the shape of the peak degrades gradually with increasing $T$, and an application of a small field ($<2$ T) only weakly affects its width, as shown in Supplementary Fig.~2(a). More straightforwardly, the ratio $l/d \sim 8$ can be estimated \cite{GramichJ-ContactSize, GallD-MeanFreePath, JangH-CeCoIn5PCS}, strongly pointing to a ballistic regime in our measurements \cite{EatonAG-NC2024, YGNaidyuk-PCSBook, AokiD-UTe2dHvA2022} (where $l$ is mean free path, and $d$ is effective microscopic contact diameter). Second, magnetic scattering may also induce conductance peaks at zero bias \cite{JApplebaum-PR1967, LShen-PR1968}. However, this is not relevant here because UTe$_2$ does not exhibit magnetic ordering at ambient pressure \cite{RanS-Science2019}. In addition, ZBCP are also observed in different samples (S2 and S3) with various contact sizes and resistances, seeing Fig.~2(e,f). The presence of stable ZBCP in various samples strongly suggests the existence of  zero-energy Andreev bound state, which further implies a high probability of gap function with odd $k_z$ component.

The results of PCS under pressure (measured with sample S1) are displayed in Fig.~3 and Supplementary Fig.~3. At low pressure (e.g. 0.2 GPa), the spectrum remains largely similar to that at ambient pressure, cf. Fig.~3(a), indicating that the SC order parameter likely remains unchanged. Remarkable differences are seen above the first critical pressure $\sim$ 0.3 GPa. As shown in Fig.~3(b), at 0.6 GPa, besides the ZBCP that is present for low pressure, dips appear on both sides of the peak at low temperature. These dips become more pronounced upon cooling, accompanied by the appearance of additional humps at slightly higher bias. As we will further explain below, the appearance of these side dips suggests that these ZBCPs are intrinsic and associated with the formation of ZABS. Further increasing pressure to 1.0 GPa, the main features of the spectrum remain the same. However, a new minor peak appears inside the dip at low temperature, forming a double-dip structure, as can be seen in Fig.~3(c). Similar behavior can also be identified at 1.2 GPa [Supplementary Fig.~3(c)]. Finally, at 1.4 GPa that is on the verge between SC and magnetically ordered phases, the central peak is substantially broadened with its top being nearly flat. Side-dip structure is also present at this pressure. To better visualize the evolution of PCS with pressure, we present contour plots of $G_\text{NS}(V,~T)$ in the bottom panels of Fig.~3. \\

\textbf{DISCUSSION}

To elucidate the evolution of the PCS of UTe$_2$ with applied pressure, we analyzed the data in two steps. First, we theoretically simulated the $G_\text{NS}(V)$ in the framework of an extended BTK model, aiming to qualitatively screen out the potential pairing symmetry. After that, numerical fittings were conducted to extract the parameters such as SC gap $\Delta$, the Dynes broadening factor $\Gamma$, and the junction barrier strength $Z$. In accordance with the experimental setup, we consider a planar NM-SC junction whose interface is perpendicular to the crystallographic $\mathbf{c}$-axis of UTe$_2$. For spin triplet SCs, the Andreev reflection is a more complex process as compared with that for conventional spin-singlet ones, in the sense that the reflection can take place through multiple channels, $\Phi_{1}=|\uparrow \uparrow\rangle$, $\Phi_{0}=(|\uparrow \downarrow\rangle + |\downarrow \uparrow \rangle)/\sqrt{2}$, and $\Phi_{-1}=|\downarrow \downarrow\rangle$. A cartoon illustration is presented in Supplementary Fig.~5.

In short, the differential conductance for PCS takes the following form,
\begin{equation}
    G_\text{NS}(V,T)=G_0\int_{-\infty}^{+\infty}\frac{\partial f(E-eV)}{\partial V}\sum_{{\mathbf{k}}_{\parallel}}\left\{1+\frac{1}{2}\sum_{\alpha,\beta}\left[|a_{\alpha\beta}(E+i\Gamma)|^2-|b_{\alpha\beta}(E+i\Gamma)|^2\right]\right\}{\rm d}E,
\end{equation}
where $G_0$ represents the conductance of the junction in the normal state; $a_{\alpha\beta}$ and $b_{\alpha\beta}$ are functions of $Z$, $\Gamma$ and $\Delta$, standing for the Andreev reflection and normal reflection coefficients, respectively; $\alpha$ and $\beta$ denote the spin states. Details of the calculation are provided in the \textbf{SI}. We first only consider the representative cases with pure $p_x$, $p_y$ and $p_z$-wave gap functions. The numerical results are displayed in Fig.~4 and Supplementary Fig.~6. In the limit of weak junction barrier (i.e. small $Z$) of our planar junction, the conductance spectra of all three states are essentially the same. The differential conductance exhibits a peak whose value is obviously less than 2 regardless of the choice of other parameters in Fig.~4(b) (i.e. $\Gamma$, $T$ and $\Delta$). This is similar to the behavior of other unconventional pairings, such as the $d_{x^2-y^2}$-wave pairing in the cases of various junction surface orientations~\cite{TanakaY-PRL1995,DagheroD-NC2012}. As $Z$ increases, the conductance of the $p_{x(y)}$- and $p_z$-wave pairings behave in fundamentally different manners. For the $p_{x(y)}$ pairing, the conductance peak diminishes with increasing $Z$ and eventually forms a gapped spectrum at large $Z$. In contrast, the peak for the $p_z$-wave sharpens as $Z$ increases, reaching a value that readily exceeds 2 at large $Z$. These pronounced ZBCPs originate from the zero-energy Andreev bound states at the junction interface (Fig.~1), and are robust against FS anisotropy (Supplementary Fig.~14) and multiplicity \cite{Daghero-PCS-anisotropy}. Their formation is closely related to the sign change of the gap function $\Delta_{\textbf{k}}$ between the incident wavevector $\textbf{k}=(k_x,k_y,k_z)$ and the reflected quasiparticles, and thus allows for localized sub-gap states to develop. No such sign change occurs for pure $p_x$ and $p_y$ pairings, and hence no bound state formation therein. Notably, bound states can still form when $p_z$ and $p_{x(y)}$ pairings mix, as long as the sign-changing $p_z$ component is finite. However, depending on the details of mixing, the sub-gap conductance may exhibit more complicated structures, as can be seen in e.g. Supplementary Fig.~7. As a side effect, at large $Z$ the presence of surface bound states tends to deplete the tunneling probability of the bulk continuum, which may then induce conductance side dips around the bias voltage $V=\pm|\Delta|$. These side dips, however, are not robust against pair-breaking scattering and may be smeared if the Dynes broadening $\Gamma$ is large, seeing below.

We then consider pairings in various IRs as proposed for UTe$_2$ (see Table 1) \cite{XuY-PRL2019, HMatsumura-JPSJ2023, FTheuss-NP2024, QGu-Science2025}, with representative results displayed in Supplementary Fig.~7. Firstly, the conductance peak of a generic $A_u$ pairing typically splits at finite $Z$ [Supplementary Fig.~7(a)], unless $\Gamma$ is sufficiently large to smear out the splitting [Supplementary Fig.~8]. Such behavior does not agree with the definitive ZBCPs seen in our samples [Fig.~2(d-f)]. Secondly, since the $B_{1u}$ pairing lacks a component that changes sign upon taking $k_z$ to $-k_z$, it cannot exhibit conductance side dips at any values of $Z$ and $\Gamma$; this situation cannot be improved even after taking into account the high-order $k_xk_yk_z$ term in $B_{1u}$ (Supplementary Fig.~13). Lastly, the $B_{2u}$ and $B_{3u}$ pairings are indistinguishable in this junction, and they both can qualitatively capture all main features of the experimental spectra in Figs.~2 and 3, and Supplementary Fig.~3.

With these in mind, we then compare and fit the experimental spectra mainly in the framework of $B_{2u}$ (or $B_{3u}$). The fittings for the ambient-pressure data for three samples are provided in Fig.~2(d-f), and the obtained parameters are listed in Supplementary Tab.~1. Although the three samples exhibit distinct profiles of PCS, a common feature is that the peak values of the $G_\text{NS}(V)$ are all less than 1.2, significantly lower than 2. This suggests that the $p_{x(y)}$ pairing component plays a more dominant role than the $p_z$ pairing. Albeit of the different values of $Z$ and $\Gamma$ obtained considering junction variation, all three samples yield a similar set of gap amplitudes $\Delta_{x(y)}$ and $\Delta_z$ with the ratio $\Delta_z/\Delta_{x(y)} \approx 0.57(5)$, which bolsters our confidence in these fittings. The temperature dependence of $\Delta_{x(y)}$ and $\Delta_z$ of sample S1 at ambient pressure is presented in Supplementary Fig.~2(c); they both strongly violate the BCS's prediction, which again demonstrates the unconventional nature of SC in UTe$_2$. We note that the gap amplitudes extracted from the BTK fittings may not necessarily coincide strictly with the thermodynamic gap values obtained from other probes; for instance, in RbCr$_3$As$_3$, another candidate of spin-triplet superconductor ($T_\text{c} \sim 7$ K), while PCS revealed SC gap as large as $\sim 2.5$ meV, the STM measurements reported only a small gap $\sim 0.8$ meV \cite{Liu-RbCr3As3}. The inherent mechanism for such discrepancy is not clear and requires further investigations. Nevertheless, since our PCS measurements are performed in the ballistic regime, these fitted amplitudes provide effective spectroscopic energy scales for characterizing the main spectral features and tracking the pressure evolution of the relative pairing components. It is also worthwhile to mention here that an earlier PCS experiments on (0~0~1) and (0.4~0.6~0.7) facets at ambient condition suggested dominant-$p_y$ wave gap function \cite{HYoon-npjQM2024}; a recent zero-energy surface state visualization by STM measurements on (0~$-1$~1) facet most favors $B_{3u}$ IR \cite{QGu-Science2025}; their conclusions are qualitatively consistent with ours for ambient pressure.

Figure 5 summarizes the fitting in different regions of the phase diagram. Representative fitting curves are displayed in Supplementary Figs.~9-11. For small pressure $p=0.2$ GPa, the results are quite similar to those at ambient pressure, and the $p_{x(y)}$ pairing is still found to dominate with $\Delta_z/\Delta_{x(y)} \approx 0.64(2)$. For $p=0.6$ GPa, however, the amplitude of ZBCP increases to $\sim$ 1.70, and most importantly, dips emerge symmetrically on the two sides of the central peak. These dips can only be reproduced by increasing the weight of $p_z$ component, as shown in Fig.~5(c). The fitting yields $\Delta_z/\Delta_{x(y)}\approx 1.03$. The situation at 0.8 GPa is similar to that at 0.6 GPa, and the amplitude of the ZBCP is further increased to 2.08, cf Fig.~5(d). At even higher pressure, the standard fitting procedure can hardly capture the double-side-dip feature (Supplementary Fig.~11). The exact origin of the double side dips remains elusive. One may conjecture that it arises from multiple gaps. An alternative possibility is due to pressure inhomogeneity, in the regard that the NM-SC junction may cover several SC domains with different gap parameters. For both cases, a linear superposition model to the total differential conductance ($G_{\text{NS}}^{\text{total}}$) are applicable:
\begin{align}
G_{\text{NS}}^{\text{total}}(V) &= \sum_i w_iG_{\text{NS}}^i(V),
\label{eq:phase_mixture_with_domain}
\end{align}
where $G_{\text{NS}}^i(V)$ is the differential conductance for each channel, and $w_i \in [0,1]$ characterizes the fractional contribution. This model can fit the multi-dip behavior rather well as shown in Fig.~5(e). Although this model does not account for the complex inter-band couplings inherent to a rigorous multi-band framework, it provides an effective estimate to the energy scale of SC order parameters \cite{SzaboP-PRL2001,DagheroD-SST2010}. The ratio between the $p_z$ and $p_{x(y)}$ components is thus defined as
\begin{align}
\frac{\langle \Delta_z \rangle}{\langle \Delta_{x(y)} \rangle} \equiv \frac{\sum_i w_i \Delta_z^i}{\sum_i w_i \Delta_{x(y)}^i},
\label{eq:ratio}
\end{align}
the pressure dependence of which at 0.3 K is presented in the inset of Fig.~5(a). It is clear that the weight of $p_z$ wave pairing increases drastically going from the SC1 to SC3, and becomes dominant in the latter region. For $p=1.2$ GPa, $\frac{\langle \Delta_z \rangle}{\langle \Delta_{x(y)} \rangle}$ reaches $\sim1.30$.

To gain insight into the transition from SC2 to SC3, we analyzed the temperature dependent PCS at fixed $p=0.8$ GPa, as shown in Fig.~5(f). Remarkably, with decreasing temperature, the ratio $\frac{\langle \Delta_z \rangle}{\langle \Delta_{x(y)} \rangle}$ first reduces, and then increases, with a minimum near 1.5 K. The results for $p=0.6$ and 1.2 GPa are supplemented in Supplementary Fig.~12, where similar temperature dependence can be observed, with minima occurring at about 1.6 and 0.7 K, respectively. In contrast, this ratio is nearly temperature-independent at $p=0$ [Supplementary Fig.~12(d)]. It is particularly interesting to note that the critical temperatures defined in association with the minimum of $\frac{\langle \Delta_z \rangle}{\langle \Delta_{x(y)} \rangle}$ match well with the boundary between SC2-SC3 that was determined through AC calorimetry and AC magnetic susceptibility \cite{ThomasSM-SciAdv2020,WuZ-PRL2025}, as shown in Fig.~5(a) (red open circles). The peculiar non-monotonic temperature dependence of $\frac{\langle \Delta_z \rangle}{\langle \Delta_{x(y)} \rangle}$ is more likely indicative of an underlying phase transition between the SC2 and SC3 states than a mere crossover.

Additional remarks in order:

(1) Is $A_{u}$ pairing possible for ambient-pressure UTe$_2$? While $B_{2u}$ (or $B_{3u}$) model provides a compelling fit to our data, we should emphasize that our results cannot entirely exclude this possibility. In fact, $A_{u}$ was supported by an ambient-pressure $^{125}$Te NMR experiment where Knight shift - proportional to the spin susceptibility - was found to decrease below $T_c$ for magnetic field along all principal axes \cite{HMatsumura-JPSJ2023}, as well as thermal conductivity measurements that pointed to a fully gapped pairing state \cite{Suetsugu-ThermalconductivitySA2024}. These works suggested that the SC order parameter $\mathbf{d}$ may contain non-negligible components along all $\mathbf{\hat{x}}$, $\mathbf{\hat{y}}$ and $\mathbf{\hat{z}}$ axes. However, we should also note that the change of Knight shift is rather anisotropic, meaning that a certain component of $\mathbf{d}$ may be rather small. If either the $\hat{\mathbf{x}}$ or $\hat{\mathbf{y}}$ component is negligible, the tunneling spectrum will be similar to those of $B_{2u}$ and $B_{3u}$ states. In this case, PCS lacks the necessary resolution to tell them apart.

(2) What can the order parameters be for pressurized UTe$_2$? From our PCS measurements, it is fairly likely that SC1-3 are mutually independent phases. First of all, SC3 differs from SC1 strikingly in that the former is $p_z$-dominant while the latter is more dominated by $p_{x(y)}$ pairing. The peculiar temperature dependence of $\frac{\langle \Delta_z \rangle}{\langle \Delta_{x(y)} \rangle}$ also hints a phase transition taking place at the boundary between SC2-SC3. It should be pointed out that an earlier $^{125}$Te NMR experiment at 1.2 GPa \cite{KinjoK-SciAdv2023} manifested that the Knight shift remains constant when it enters the SC2 phase from the normal state, but starts to drop at the boundary of SC2-SC3. This indicates that the $\mathbf{\hat{y}}$-component of $\mathbf{d}$ should be vanishingly small for SC2 but considerable for SC3. In this sense, and also taking into account our PCS experiment, $B_{2u}$ appears the most promising for SC2. Below the SC2-SC3 boundary, a secondary SC order parameter starts to condense, probably yielding a superposition of $B_{2u}$ and $B_{3u}$ pairings in SC3. If so, the transition at SC2-SC3 would be accompanied with additional symmetry breaking. However, we must admit that before a more systematic PCS investigation is made on other surfaces of UTe$_2$, it is premature to draw a precise conclusion.

(3) Does a quantum critical region exist in UTe$_2$ under pressure? Notably, the conductance spectrum near 1.4 GPa exhibits an anomalous broadening with a nearly flat top, making it challenging to adequately describe within a simple single-component BTK framework. This drastic spectral evolution may suggest an underlying change in the system's ground state, rather than a mere enlargement of the superconducting gap. Given that previous pressure studies have reported a possible magnetic quantum critical point (QCP) nearby \cite{ThomasSM-SciAdv2020}, we speculate that the anomalously broadened spectrum could be related to this criticality. Specifically, the proliferated critical fluctuations can lead to a significant enhancement of scattering rate, the latter of which can increase the Dynes broadening factor $\Gamma$. While this scenario offers a plausible explanation for the 1.4 GPa data, we cautiously note that further experimental evidence is required to draw a definitive conclusion regarding its connection to a QCP.

In summary, we systematically investigated the pressure-evolution of point-contact spectroscopy in UTe$_{2}$. Clear and reproducible ZBCP is observed on the (0~0~1) surface, and the profile of the ZBCP is strongly pressure dependent with pronounced side-dip structure appearing for pressures above 0.3 GPa, likely indicating multiple SC phases whose pairing gap functions contain $k_z$-odd components. Quantitative analysis based on an extended BTK model suggests $B_{2u}$ (or $B_{3u}$) as the most conceivable irreducible representation for both ambient and pressurized cases, with the ratio $\frac{\langle \Delta_z \rangle}{\langle \Delta_{x(y)} \rangle}$ serving as a useful parameter to differentiate the SC1-3 phases. It is found that the weight of the $p_z$-wave component increases drastically from SC1 to SC3, and the SC2-SC3 transition may involve the emergence of an additional superconducting order-parameter component. Our work not only provides spectroscopic evidence for the existence of multiple SC phases in pressurized UTe$_2$, but also places critical constraints on its superconducting order parameter, thus collectively establishing UTe$_2$ as a prototypical system for studying intertwined electronic orders upon pressure tuning. \\

\textbf{METHODS}\\

\textbf{Single crystals preparation and characterization}

High-quality single crystals of UTe$_{2}$ were grown by a molten soft flux method as described elsewhere \cite{SakaiH-PRM2022,EatonAG-NC2024}. We employed a mixture of NaCl and KCl as a flux. High-purity U (99.999\%), Te (99.9999\%), NaCl (99.99\%) and KCl (99.999\%) were mixed in a molar ratio of 1:1.71:60:60. To prevent contamination or oxidation of the starting materials, the entire process was carried out in an Ar-filled glove box. The mixture was transferred into a carbon crucible and sealed in quartz tube which was then heated up to 473 K and dwelt for 12 hours. The temperature was then raised to 723 K in 24 hours, held at this temperature for an additional 24 hours, and subsequently increased to 1223 K at a rate of 0.35 K/min and maintained for another 24 hours. After that, the temperature was gradually reduced to 923 K at a rate of 0.03 K/min, kept stable for 24 hours, and finally cooled to room temperature over the next 24 hours. This controlled thermal profile ensured the synthesis of high-quality crystalline materials.  \\

\textbf{Measurements under pressure}

Electrical transport measurements of UTe$_2$ (sample S1) were performed in an 8 T superconducting magnet system, using a $^3$He refrigerator (Oxford Instruments). Prior to junction fabrication, the surfaces of the sample were fine-polished to obtain a fresh crystallographic (0~0~1) plane. A soft point-contact junction was then prepared by depositing a minimal droplet of Au epoxy (54L-2210) to glue the gold wires onto the fresh surface. The small Au particles in the epoxy are solutes in the organic suspension, enabling the construction of a point contact with a small effective contact area. The Au epoxy was heated at 423 K for 30 minutes before the sample was transferred into a pressure cell. All these processes were conducted in an argon-filled glove box. A hybrid piston-clamp type cell was used to achieve quasi-hydrostatic pressures up to $\sim$ 1.8 GPa with Daphne 7373 oil serving as the pressure-transmitting medium. This rigorous sample preparation protocol preserves the intrinsic surface integrity of UTe$_{2}$ and provides a stable electrical interface that is essential for the accurate probing of quasi-particle dynamics. The pressure sequence was as follows: $0 \rightarrow 0.2 \rightarrow 0.6 \rightarrow 1.0 \rightarrow 1.4 \rightarrow 0.3 \rightarrow 0.8 \rightarrow 1.2~\mathrm{GPa}$. To check the reproducibility of the results, two additional samples (S2 and S3) were measured at ambient pressure, and pressure-cycling test was performed on sample S1. \\

\textbf{Data availability}

All data that support the findings of this study are available within the paper and the Supplementary Information. Raw data generated during the current study are available from the corresponding authors upon request. \\



\emph{}\\
\textbf{ACKNOWLEDGEMENTS}\\

A portion of this work was carried out at the Synergetic Extreme Condition User Facility (SECUF, https://cstr.cn/31123.02.SECUF). A.G.E. acknowledges support from the Henry Royce Institute for Advanced Materials through the Equipment Access Scheme enabling access to the Advanced Materials Characterization Suite at Cambridge (EP/P024947/1, EP/M000524/1, and EP/R00661X/1).\\

\textbf{FUNDING STATEMENT}\\

This work is supported by the National Key R\&D Program of China (2023YFA1609600, 2022YFA1602602, and 2022YFA1402200), National Natural Science Foundation of China (U23A20580, 12374042, 12574147, 52588101, W2511006 and 12494592), Beijing National Laboratory for Condensed Matter Physics (2024BNLCMPKF004), and the New Cornerstone Science Foundation (Grants No. NCI202509). W.H. acknowledges support from a startup fund at Great Bay University and the Dongguan Key Laboratory of Artificial Intelligence Design for Advanced Materials.  \\

\textbf{AUTHOR CONTRIBUTIONS}

Y.L., X.L. and H.Y. conceived and designed the experiments; Z.W., A.C., M.V., N.L., B.Z., A.G.E. and C.Z. grew and characterized high-quality single crystalline samples; S.Z., F.S. performed most of the experiments with the aids from Y.Z., W.Q., Z.W., H.Z., Y.M., Z.Y., K.Y., J.Z., Y.Z., K.L., G.L. and J.L; S.Z. and Z.Q. carried out numerical simulations and fittings with the aids from J.L.Z. and W.H; S.Z., Z.Q., W.H., X.L., H.Y. and Y.L. discussed the data, interpreted the results, and wrote the paper with input from all the authors. \\

\textbf{COMPETING INTERESTS}

The authors declare no competing interests.\\

\textbf{ADDITIONAL INFORMATION}\\
\textbf{Supplementary information} The online version contains supplementary material available at ******** \\

\textbf{Correspondence} and requests for materials should be addressed to Wen Huang, Huiqiu Yuan, Xin Lu or Yongkang Luo.\\

\textbf{Reprints and permission information} is available at \\
http:$\backslash\backslash$www.nature.com/reprints \\

\textbf{Publisher's note} Springer Nature remains neutral with regard to jurisdictional claims in published maps and institutional affiliations.

\newpage
\begin{table*}[!htp]
\caption{\label{Tab1} Possible pairing symmetries and the corresponding forms of the gap functions of UTe$_2$ classified according to the D$_{2h}$ point group~\cite{XuY-PRL2019, HMatsumura-JPSJ2023, FTheuss-NP2024, QGu-Science2025}. Note that we here ignore high-order terms beyond linear order of $k_i$ ($i=x,y,z$), seeing more details in Supplementary Fig.~13. Zero-energy Andreev bound state (ZABS) are expected on the (0~0~1) surface for $B_{2u}$ or $B_{3u}$ representation.}
\begin{ruledtabular}
\begin{tabular}{cccc}
                       \multicolumn{4}{c}{UTe$_2$ D$_{2h}$}                                                                        \\ \hline
      IR          & Basis functions                                                                            &  Spin component     &  ZABS on $(0~0~1)$?   \\ \hline
      $A_{u}$     & $\Delta_x k_x\hat{\mathbf{x}}+\Delta_y k_y\hat{\mathbf{y}}+\Delta_z k_z\hat{\mathbf{z}}$   &                     &      No              \\
      $B_{1u}$    &  $\Delta_y k_y\hat{\mathbf{x}}+\Delta_x k_x\hat{\mathbf{y}}$                               & $\hat{\mathbf{z}}$  &      No              \\
      $B_{2u}$    & $\Delta_z k_z\hat{\mathbf{x}}+\Delta_x k_x\hat{\mathbf{z}}$                                & $\hat{\mathbf{y}}$  &     Yes              \\
      $B_{3u}$    & $\Delta_z k_z\hat{\mathbf{y}}+\Delta_y k_y\hat{\mathbf{z}}$                                & $\hat{\mathbf{x}}$  &     Yes              \\
\end{tabular}
\end{ruledtabular}
\small
\vspace*{-10pt}
\begin{flushleft}
\end{flushleft}
\normalsize
\end{table*}

\newpage
\begin{figure*}[!htp]
\hspace*{-0pt}
\vspace*{-0pt}
\includegraphics[width=16.5cm]{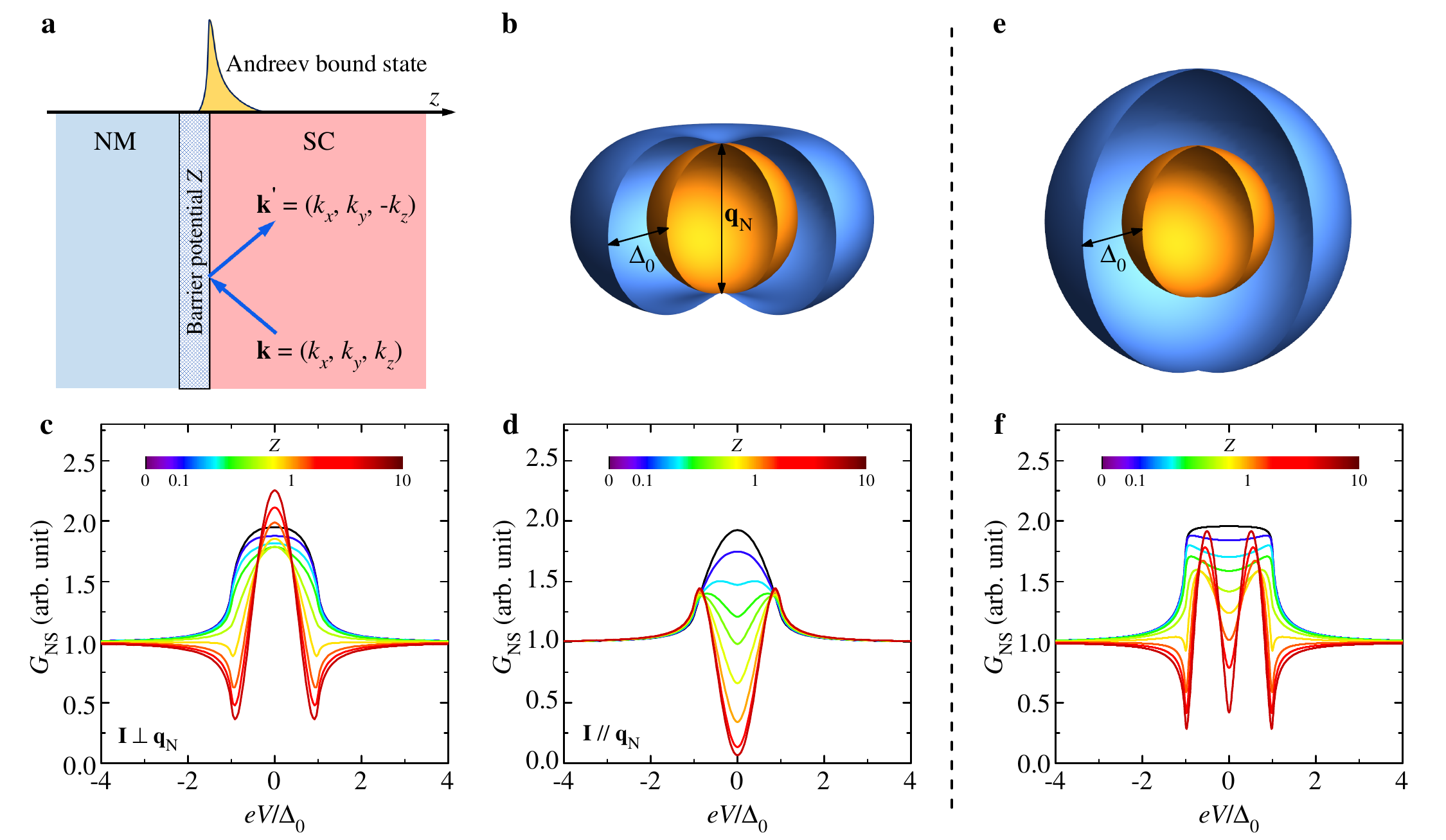}
\label{Fig1}
\end{figure*}
\vspace*{-20pt}
\begin{spacing}{1.40}
\small{\textbf{Figure 1 $|$ Gap structure and Andreev bound state (ABS) in unconventional superconductors.} \textbf{a} A cartoon illustration of the reflection of Cooper pairs at the boundary (characterized by a barrier potential $Z$) of normal metal (NM) and superconductor (SC) that changes the scattering wavevector $\mathbf{k}=(k_x,~k_y,~k_z)$ into $\mathbf{k}'=(k_x,~k_y,-k_z)$. If certain component of the SC gap function [$\Delta(\mathbf{k})$] changes sign upon taking $k_z$ to $-k_z$, an ABS can be trapped near the boundary. \textbf{b} Nodal gap structure with maximum $\Delta_0$. The inter-nodal scattering is denoted by the wavevector $\mathbf{q}_\text{N}$. \textbf{c} For tunneling current $\mathbf{I}\perp\mathbf{q_\text{N}}$, the sign change in $\Delta(k_z)$ $\rightarrow$ $\Delta(-k_z)$ leads to the formation of zero-energy ABS (ZABS) and hence a sharp zero-bias conductance peak (ZBCP) in the differential conductance ($G_\text{NS}$) that readily exceeds 2 in the large-$Z$ limit. \textbf{d} Such a ZABS is absent if the tunneling current is along $\mathbf{q}_\text{N}$. For an arbitrary $\mathbf{I}$ with a finite component perpendicular to $\mathbf{q_\text{N}}$, ZABS persists, albeit with reduced intensity. \textbf{e} Sketch of a fully-gapped structure, e.g. $A_u$ representation. \textbf{f} $G_\text{NS}$ does not exhibit a ZABS in this case; instead, a pronounced dip is expected at zero bias at large $Z$.} \\
\end{spacing}

\newpage
\begin{figure*}[!htp]
\vspace*{0pt}
\hspace*{-0pt}
\includegraphics[width=16.5cm]{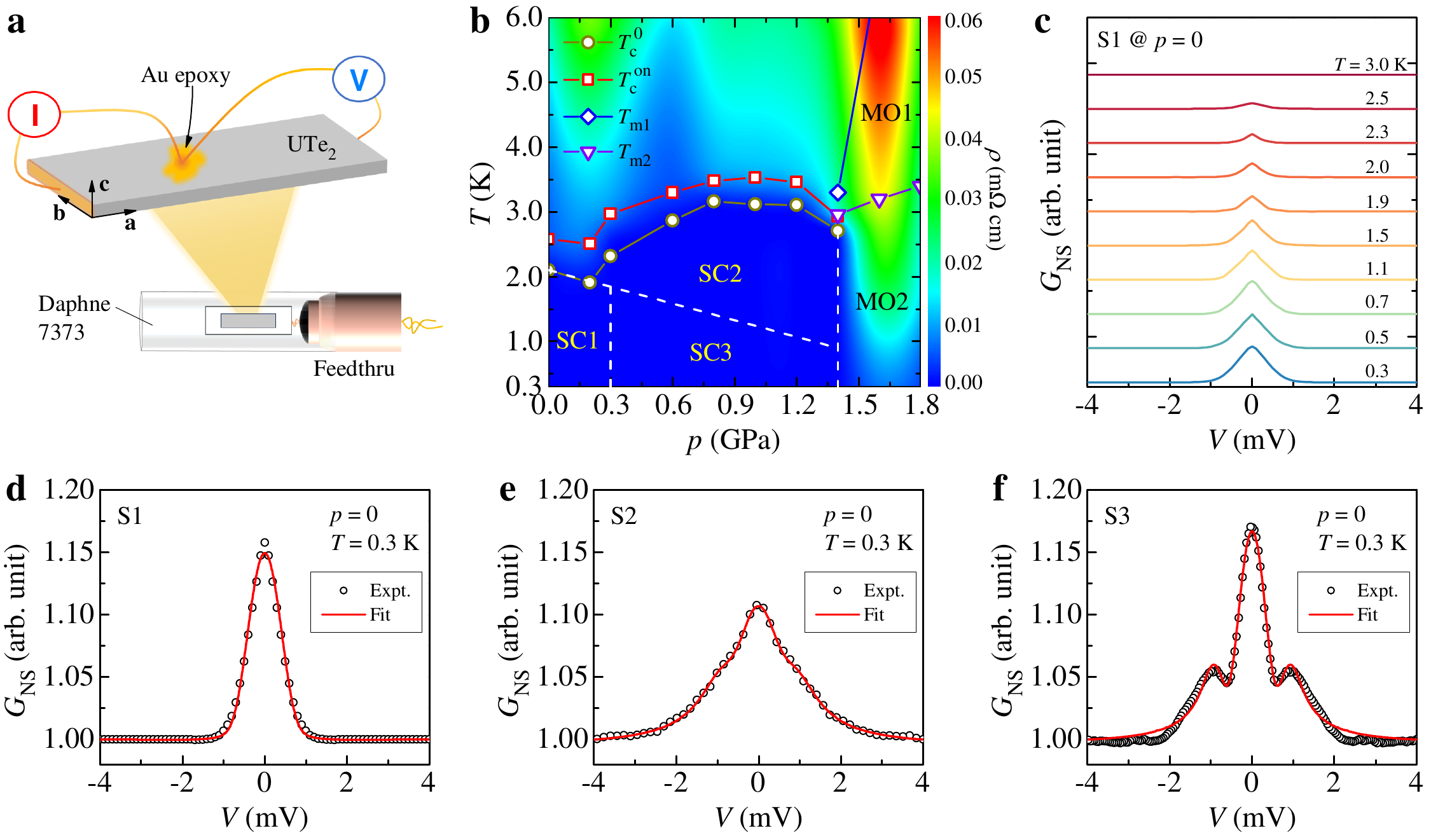}
\label{Fig2}
\end{figure*}
\vspace*{-20pt}
\begin{spacing}{1.40}
\small{\textbf{Figure 2 $|$ Point-contact spectroscopy (PCS) of UTe$_2$ at ambient pressure.} \textbf{a} A sketch of point contact spectroscopy measurement under pressure. The junction is made on the (0~0~1) plane. \textbf{b} Phase diagram of UTe$_2$ constructed by a contour plot of $\rho(p, T)$. The abbreviations are: SC superconductor, MO magnetic order. The dashed lines are reproduced from previous specific heat and magnetic susceptibility studies \cite{AokiD-JPSJ2020,ThomasSM-SciAdv2020,WuZ-PRL2025}. These dashed lines divide the SC region into three possible phases, SC1-3. \textbf{c} The differential conductance ($G_\text{NS}$) of sample S1 at various temperatures. The curves are shifted vertically for clarity.
\textbf{d}-\textbf{f} $G_\text{NS}$ at 0.3 K for samples S1, S2 and S3, respectively. The red lines are fittings of the experimental data (open circles) to the extended BTK model in $B_{2u}$ (or $B_{3u}$) representation. The obtained fitting parameters are summarized in Supplementary Table 1. }\\
\end{spacing}

\newpage
\begin{figure*}[!htp]
\vspace*{-0pt}
\hspace*{-10pt}
\includegraphics[width=17cm]{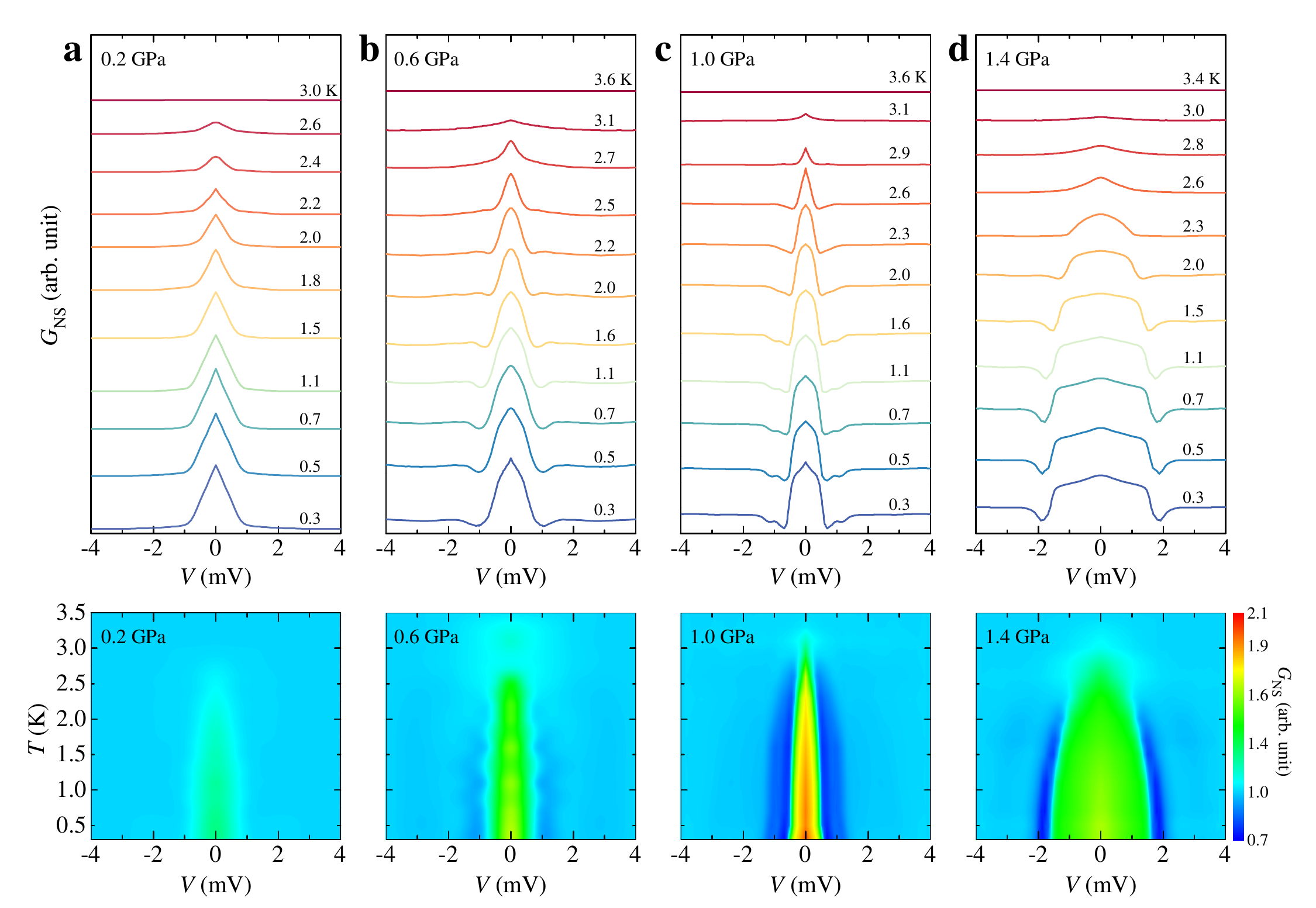}
\label{Fig3}
\end{figure*}
\vspace*{-30pt}
\begin{spacing}{1.40}
\small{\textbf{Figure 3 $|$ Point-contact spectroscopy (PCS) of UTe$_2$ under pressure.} Upper panels, differential conductance ($G_\text{NS}$) at selected pressures, \textbf{a} 0.2 GPa; \textbf{b} 0.6 GPa; \textbf{c} 1.0 GPa; and \textbf{d} 1.4 GPa. The lower panels display the false-color contour plots of $G_\text{NS}(V,T)$. } \\
\end{spacing}

\newpage
\begin{figure*}[!htp]
\vspace*{-0pt}
\hspace*{-0pt}
\includegraphics[width=15cm]{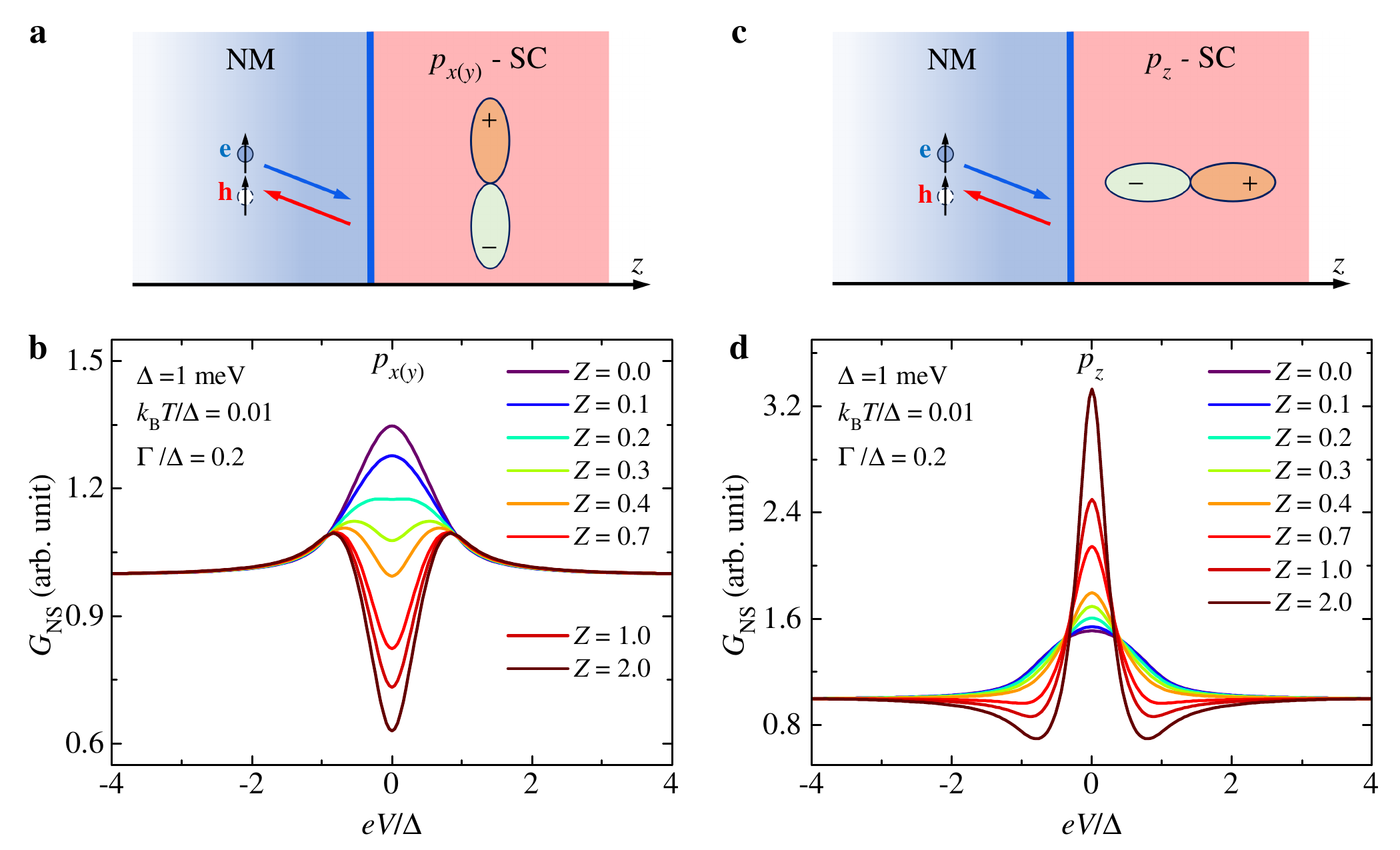}
\label{Fig4}
\end{figure*}
\vspace*{-10pt}
\begin{spacing}{1.40}
\small{\textbf{Figure 4 $|$ Simulation of Point-contact Andreev reflection spectroscopy based on the extended Blonder-Tinkham-Klapwijk (BTK) model.} The abbreviations are: NM normal metal, SC superconductor. \textbf{a-b} $p_{x(y)}$ pairing symmetry. \textbf{c-d} $p_{z}$ pairing symmetry. The simulations were performed with $\Delta=1$ meV, $k_BT/\Delta=0.01$, $\Gamma/\Delta=0.2$, and variable $Z$ listed in the figures. }    \\
\end{spacing}

\newpage
\begin{figure*}[!htp]
\vspace*{-0pt}
\hspace*{0pt}
\includegraphics[width=16.5cm]{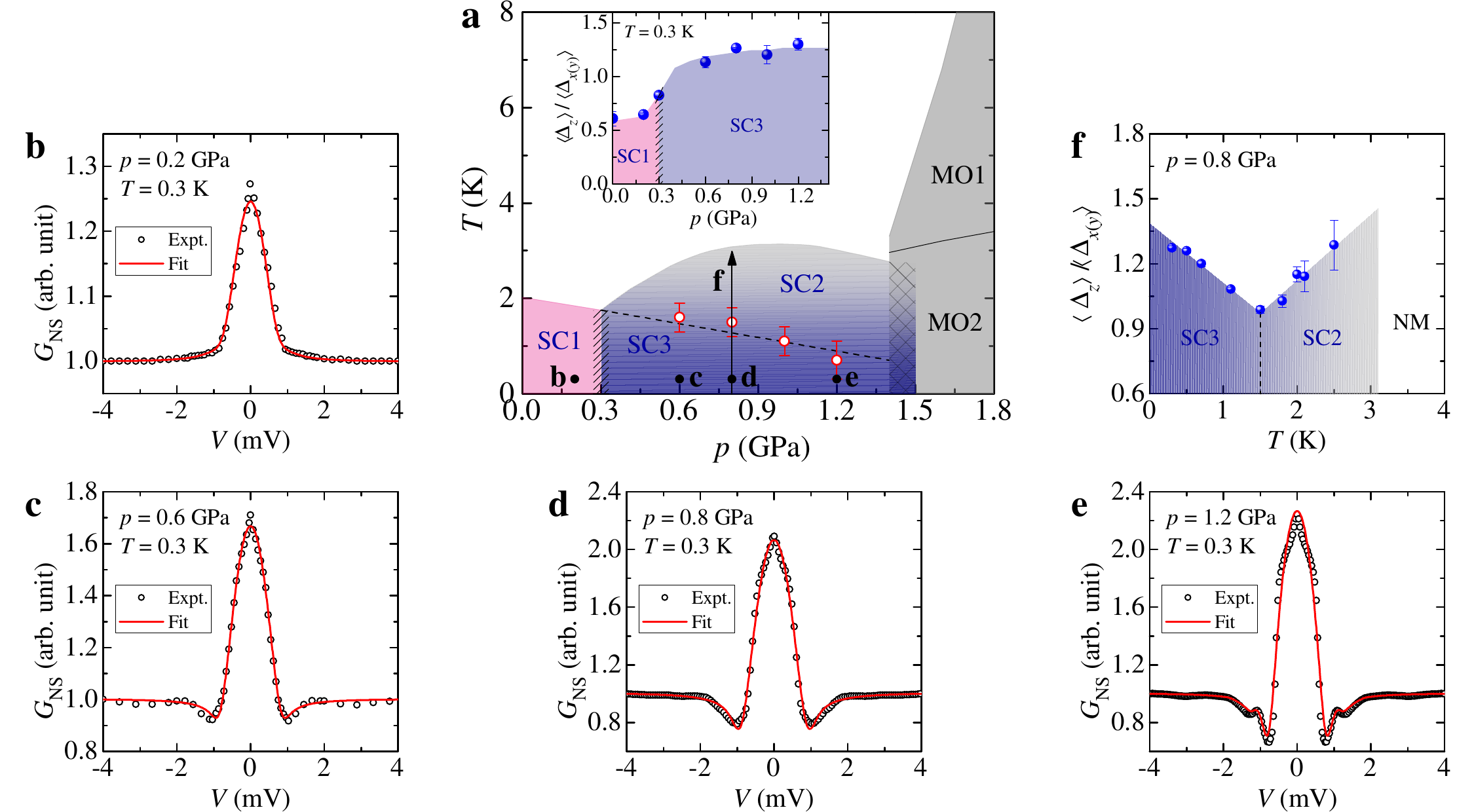}
\label{Fig5}
\end{figure*}
\vspace*{-20pt}
\begin{spacing}{1.40}
\small{\textbf{Figure 5 $|$ Blonder-Tinkham-Klapwijk fittings of the Point-contact spectroscopy (PCS) in different regions.} \textbf{a} Schematic phase diagram of UTe$_2$ under pressure. The abbreviations are: NM normal metal, SC superconductor, MO magnetic order. The boundaries of the phase diagram are generally reproduced from Fig. 2(b). The dashed line is a guide to eyes to mark the transition between SC2 and SC3 determined by earlier AC calorimetry and AC magnetic susceptibility measurements \cite{ThomasSM-SciAdv2020,WuZ-PRL2025}. The red open circles represent the critical temperatures defined as the minimum of iso-pressure temperature dependent $\langle \Delta_z\rangle / \langle \Delta_{x(y)}\rangle$, as exemplified by the results at 0.8 GPa in panel \textbf{f}. The inset presents $\langle \Delta_z\rangle/\langle \Delta_{x(y)}\rangle$ at 0.3 K as a function of $p$, which reveals an abrupt increase in the weight of $p_z$-wave pairing from SC1 to SC3. \textbf{b} Fitting for $p=0.2$ GPa and $T=0.3$ K. \textbf{c} Fitting for $p=0.6$ GPa and $T=0.3$ K. \textbf{d} Fitting for $p=0.8$ GPa and $T=0.3$ K. \textbf{e} Fitting for $p=1.2$ GPa and $T=0.3$ K. \textbf{f} $\langle \Delta_{z}\rangle/\langle\Delta_{x(y)}\rangle$ as a function of $T$ at fixed pressure $p=0.8$ GPa. } \\
\end{spacing}

\clearpage

\renewcommand{\thefigure}{S\arabic{figure}}
\renewcommand{\thetable}{S\arabic{table}}
\renewcommand{\theequation}{S\arabic{equation}}
\setcounter{table}{0}
\setcounter{figure}{0}
\setcounter{equation}{0}
\setcounter{page}{1}

\vspace{-15pt}

\begin{center}
\large
\textbf{Supplementary Information: } \\
\textbf{Multiple superconducting phases and order-parameter evolution in pressurized UTe$_2$}\\

\small
\emph{}\\
Shuo Zou$^{1*}$, Fengrui Shi$^{2*}$, Zhuolun Qiu$^{1*}$, Jia-Long Zhang$^{3}$, Yan Zhang$^2$, Weilong Qiu$^{2}$, Zhuo Wang$^{1}$, Hai Zeng$^{1}$, Yinina Ma$^{4}$, Zheyu Wu$^{5}$, Andrej Cabala$^{6}$, Michal Vali\v{s}ka$^{6}$, Ning Li$^{7}$, Zihan Yang$^{2}$, Kaixin Ye$^{2}$, Jiawen Zhang$^{2}$, Yanan Zhang$^{2}$, Kangjian Luo$^{1}$, Binbin Zhang$^{7}$, Alexander G. Eaton$^{5}$, Chaofan Zhang$^{7}$, Gang Li$^{4}$, Jianlin Luo$^{4}$, Wen Huang$^{3\dag}$, Huiqiu Yuan$^{2,8,9\ddag}$, Xin Lu$^{2,8,9\S}$, and Yongkang Luo$^{1,10\P}$ \\

$^1$ {\it Wuhan National High Magnetic Field Center and School of Physics, Huazhong University of Science and Technology, Wuhan 430074, China;}\\
$^2$ {\it Center for Correlated Matter and School of Physics, Zhejiang University, Hangzhou 310058, China;}\\
$^3$ {\it School of Physical Sciences, Great Bay University, Dongguan 523000, China;}\\
$^4$ {\it Beijing National Laboratory for Condensed Matter Physics, Institute of Physics, Chinese Academy of Sciences, Beijing 100190, China;}\\
$^5$ {\it Cavendish Laboratory, University of Cambridge, Cambridge CB3 0HE, United Kingdom;}\\
$^6$ {\it Charles University, Faculty of Mathematics and Physics, Department of Condensed Matter Physics, Prague 2 121 16, Czech Republic;}\\
$^7$ {\it College of Advanced Interdisciplinary Studies and Nanhu Laser Laboratory, National University of Defense Technology, Changsha, Hunan 410073, China;}\\
$^8$ {\it Institute of Fundamental and Transdisciplinary Research, Zhejiang University, Hangzhou 310058, China;}\\
$^9$ {\it Collaborative Innovation Center of Advanced Microstructures, Nanjing 210093, China;}\\
$^{10}$ {\it State Key Laboratory of Advanced Electromagnetic Technology, Huazhong University of Science and Technology, Wuhan 430074, China.}\\

\end{center}

\normalsize

In this \textbf{Supplementary Information (SI)}, we provide more results that will further support the discussions and conclusion in the main text, including additional resistivity and point-contact spectroscopy (PCS) data under pressure, simulations of Andreev PCS and fittings to the experimental data. \\


\textbf{SI \Rmnum{1}: Resistivity results of UTe$_2$ under pressure}\\

\begin{figure*}[!htp]
	\vspace*{-10pt}
	\includegraphics[width=16.5cm]{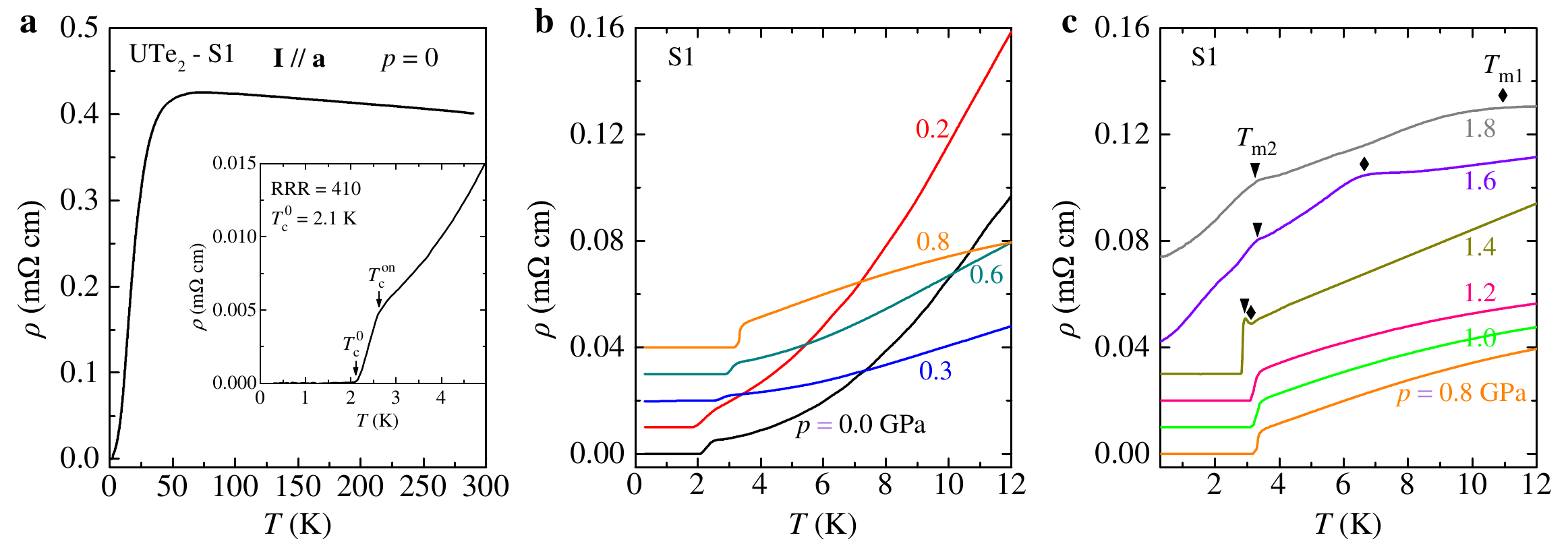}
	\label{Fig.S1}
\vspace*{-15pt}
\small
\begin{flushleft}
\justifying{
    \textbf{Supplementary Figure 1 $|$ Resistivity data of UTe$_2$ (sample S1) under pressure.} Measured with electrical current $\mathbf{I}\parallel\mathbf{a}$. \textbf{a} Main frame, electrical resistivity ($\rho$) as a function of $T$ at ambient pressure. Inset, the zoom-in plot to show the superconducting transition at low temperature. The sample is of good quality with residual resistance ratio (RRR) $\sim 410$, and SC transition temperatures $T_\text{c}^\text{on}=2.6$ K and $T_\text{c}^0=2.1$ K. \textbf{b} and \textbf{c} $\rho(T)$ profiles for under pressure. The curves are vertically shifted for clarity. The black triangles and diamonds denote the characteristic temperatures ($T_\text{m1}$ and $T_\text{m2}$) of the two magnetically ordered phases MO1 and MO2, respectively.
    }
\end{flushleft}
\normalsize
\end{figure*}

\newpage

\textbf{SI \Rmnum{2}: Additional PCS results at ambient pressure}\\

\begin{figure*}[!htp]
	\vspace*{-10pt}
	\includegraphics[width=10cm]{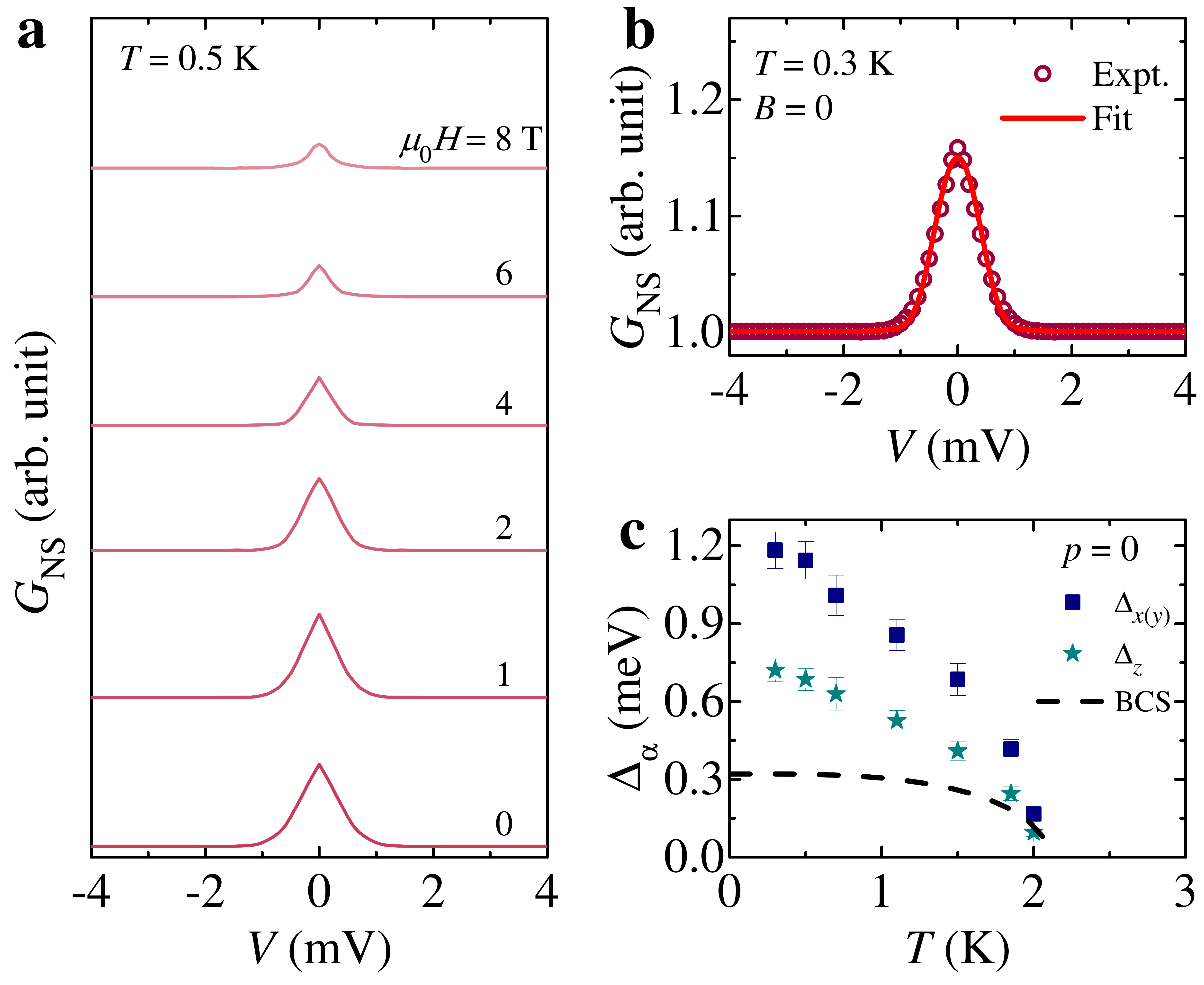}
	\label{Fig.S2}
\vspace*{-10pt}
\small
\begin{flushleft}
\justifying{
    \textbf{Supplementary Figure 2 $|$ Additional Point-contact Andreev reflection spectroscopy of UTe$_2$ at ambient pressure.} \textbf{a} Isothermal $G_\text{NS}(V)$ under different magnetic field for fixed temperature $T=0.5$ K. \textbf{b} Representative fitting of the $\text{d}I/\text{d}V$ spectrum to the extended Blonder-Tinkham-Klapwijk (BTK) model, taking $T=0.3$ K as an example. \textbf{c} The obtained gap parameters $\Delta_{x(y)}$ and $\Delta_z$ as functions of $T$. They apparently deviates from the conventional Bardeen-Cooper-Schrieffer (BCS) behavior, strongly suggesting unconventional superconductivity.
    }
\end{flushleft}
\normalsize
\end{figure*}

\newpage

\textbf{SI \Rmnum{3}: Additional PCS results under pressure}\\

\begin{figure*}[!htp]
	\vspace*{-20pt}
	\includegraphics[width=16.5cm]{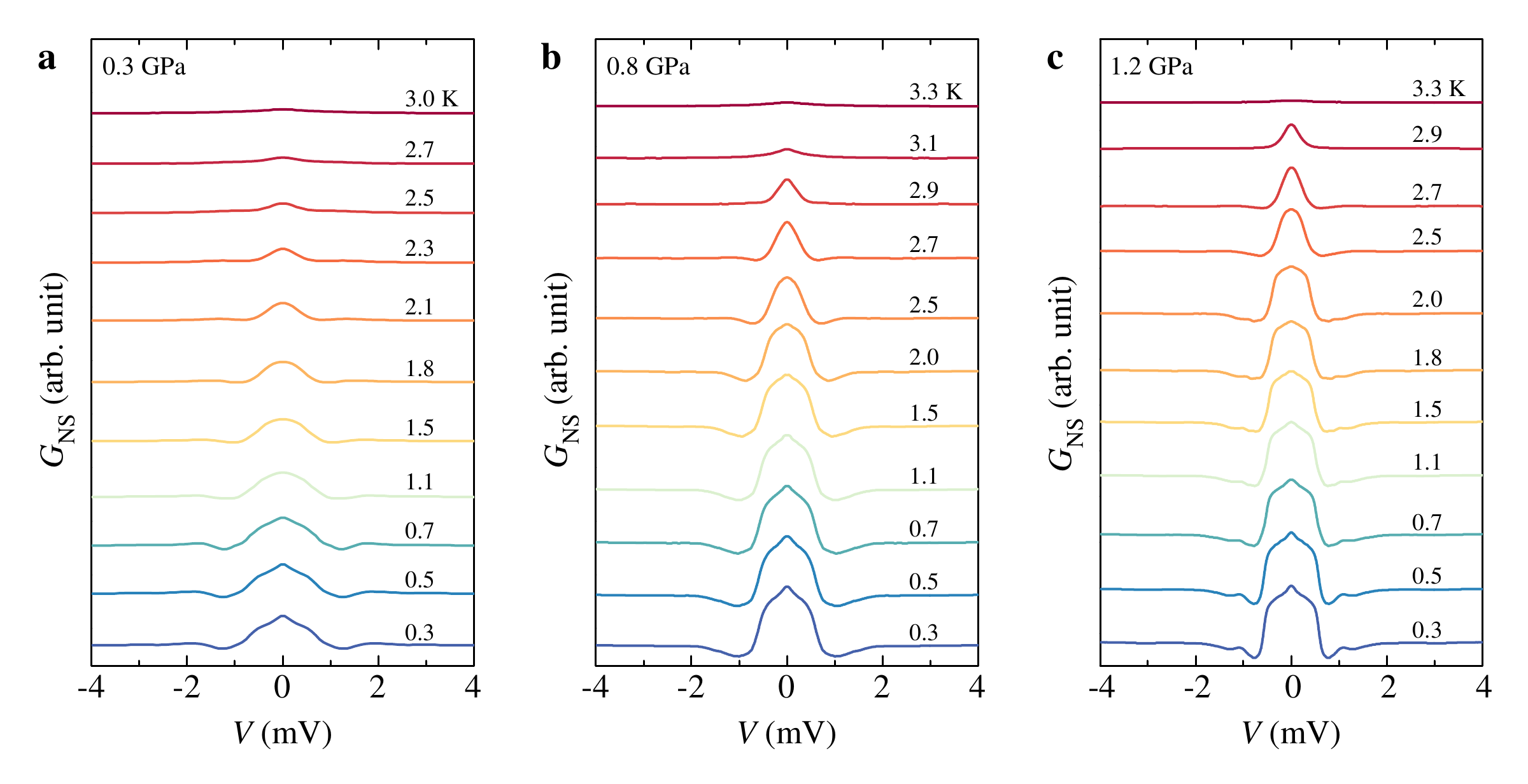}
	\label{Fig.S3}
\vspace*{-25pt}
\small
\begin{flushleft}
\justifying{
    \textbf{Supplementary Figure 3 $|$ Additional Point-contact Andreev reflection spectroscopy data of UTe$_2$ under pressure.} The differential conductance ($G_\text{NS}$) at various temperatures. \textbf{a} 0.3 GPa. \textbf{b} 0.8 GPa. \textbf{c} 1.2 GPa. The curves are shifted vertically for clarity. Note that these data were collected after releasing 1.4 GPa.
    }
\end{flushleft}
\normalsize
\end{figure*}

\begin{figure*}[!htp]
	\vspace*{-20pt}
	\includegraphics[width=16.5cm]{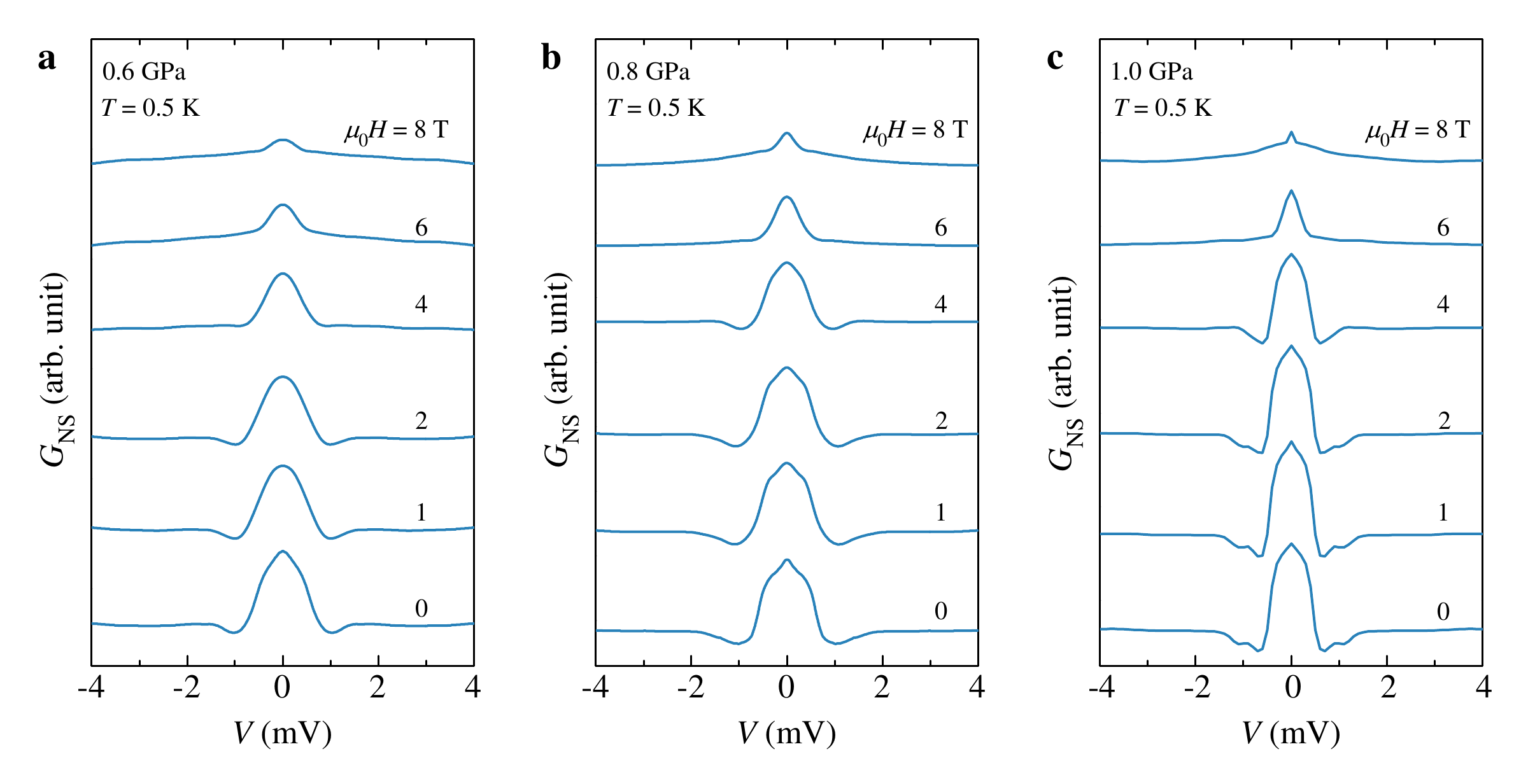}
	\label{Fig.S4}
\vspace*{-20pt}
\small
\begin{flushleft}
\justifying{
   \textbf{Supplementary Figure 4 $|$ Isothermal $G_\text{NS}(V)$ under different magnetic field for various pressures.} \textbf{a} 0.6 GPa. \textbf{b} 0.8 GPa. \textbf{c} 1.0 GPa. The curves are shifted vertically for clarity.
    }
\end{flushleft}
\normalsize
\end{figure*}

\newpage

\textbf{SI \Rmnum{4}: Extended BTK model for spin-triplet superconductivity}\\

In this section, we provide the details to deduce the expression of differential conductance ($G_\text{NS}=\text{d}I/\text{d}V$) of Andreev reflection PCS for a junction between normal metal and superconductor (NM-SC), as sketched in Supplementary Fig.~5(a). Andreev reflection refers to a scattering process that an electron incident on the interface from the NM is retro-reflected as a hole by creating a Cooper pair in the SC \cite{Andreevreflection64}. For spin-triplet SCs, the process of Andreev reflection is more complex compared with that of conventional spin-singlet counterparts due to the increase of the reflection channels. For instance, if the incident particle is a spin-up electron, the reflected particle could be either a spin-up hole through the channel of equal-spin Cooper pair $\Phi_{+1}=|\uparrow \uparrow \rangle$, or a spin-down hole through the channel of opposite-spin Cooper pair $\Phi_{0}=(|\uparrow \downarrow\rangle + |\downarrow \uparrow \rangle)/\sqrt{2}$, as depicted in Supplementary Fig.~5(b). The Blonder-Tinkham-Klapwijk (BTK) model provides a theoretical framework for analyzing the Andreev reflection PCS \cite{BTKmodel82}, which is based on the Bogoliubov-de Gennes (BdG) method.

\begin{figure*}[!htp]
	\vspace*{-0pt}
	\includegraphics[width=16cm]{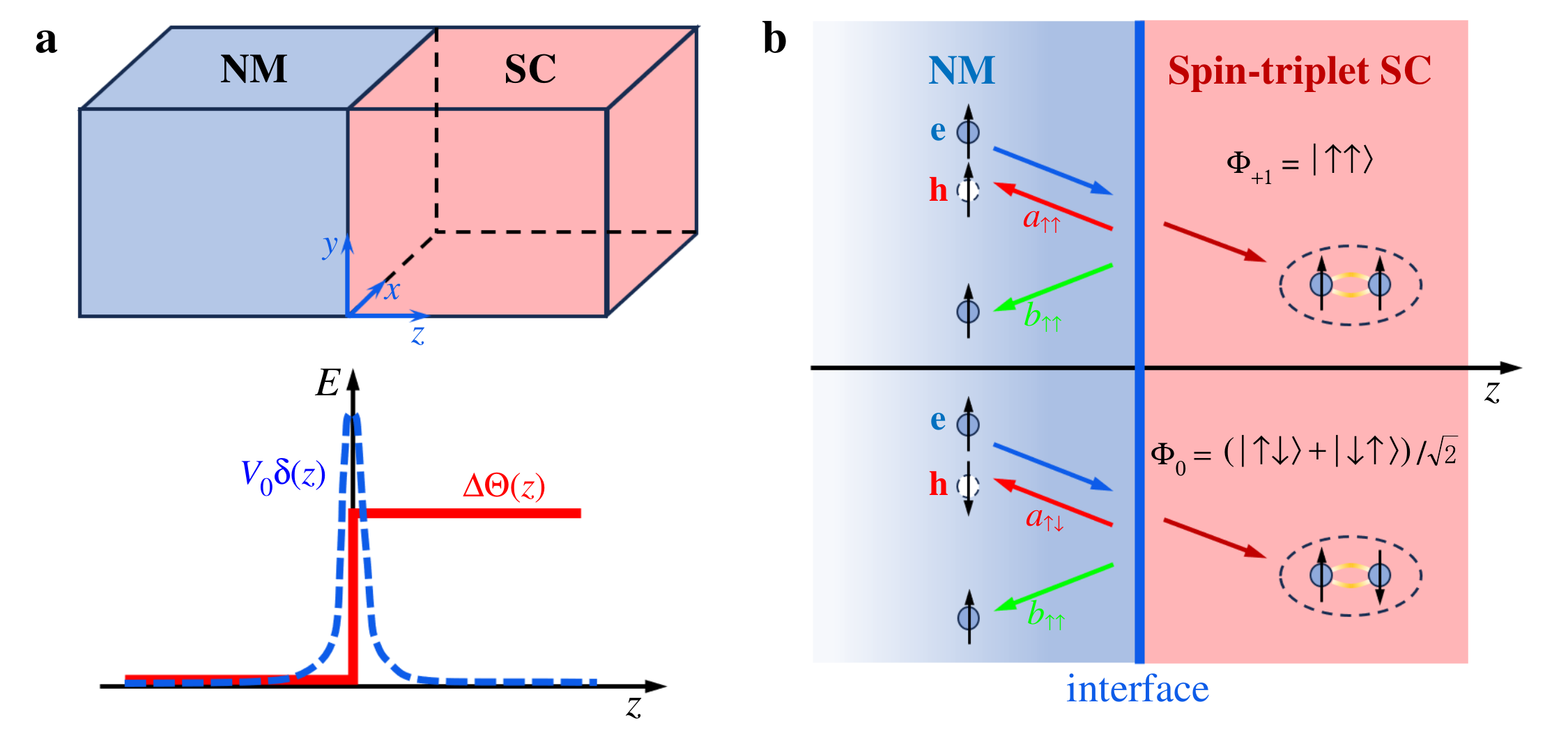}
	\label{Fig.S5}
\vspace*{-10pt}
\small
\begin{flushleft}
\justifying{
    \textbf{Supplementary Figure 5 $|$ Schematics of Andreev reflection at the interface of normal metal (NM) and superconductor (SC).}  \textbf{a} Sketch of the NM-SC junction. In this work, the junction is made by Au-UTe$_2$, on the $\mathbf{x}$-$\mathbf{y}$ crystallographic plane of UTe$_2$. \textbf{b} Andreev reflection (red arrows, $a_{\alpha\beta}$) for spin-triplet SC. For an electron (blue circle) incident at the interface, a hole (open circle) is retro-reflected by creating a Cooper pair in the SC. Note that the spin state of the reflected hole relies on the specific configuration of the Cooper pair. The green arrows denote the normal specular reflection ($b_{\alpha\beta}$) that preserves spin degree of freedom.
    }
\end{flushleft}
\normalsize
\end{figure*}

In this work, the NM-SC junction lies in the $\mathbf{x}$-$\mathbf{y}$ plane (or $\mathbf{a}$-$\mathbf{b}$ crystallographic plane), with a potential barrier $V_0\delta(z)$ at the interface $z=0$. The NM sits at the region $z<0$, while the spin-triplet SC resides in $z>0$.

The BdG equation that governs the system is given by \cite{superconductivity91}
\begin{equation}
    \check{H}_\text{BTK}({\mathbf{r}})\Phi({\mathbf{r}})=E\Phi({\mathbf{r}}),
\end{equation}
where $\check{H}_\text{BTK}(\mathbf{r})=\begin{bmatrix}
    \xi(\mathbf{r}) \hat{\sigma}_0 & \hat{\Delta}(\mathbf{r}) \\
    \hat{\Delta}^{\dagger}(\mathbf{r}) & -\xi(\mathbf{r})\hat{\sigma}_0
\end{bmatrix}$, $\xi(\mathbf{r})=-\frac{\hbar^2\nabla^2}{2m}+V_0\delta(z)-\varepsilon_F$, $\hat\Delta(\mathbf{r})=\hat\Delta e^{i\varphi}\Theta(z)$. $\varepsilon_F$ is the Fermi energy of the system, and $\hat{\sigma}_0$ is the $2\times2$ identity matrix. The gap matrix for a spin-triplet SC is defined as $\hat{\Delta}({\mathbf{k}})=i {\mathbf{d}}({\mathbf{k}})\cdot\hat{\boldsymbol{\sigma}} \hat{\sigma}_2$, where ${\mathbf{d}}=(d_1,~d_2,~d_3)$ is the vector that describes the pairing states of electrons, and $\hat{\boldsymbol{\sigma}}=(\hat{\sigma}_1,\hat{\sigma}_2,\hat{\sigma}_3)$ with $\hat{\sigma}_{i}~(i=1,2,3)$ being the Pauli matrix. We here assume ${\mathbf{d}}\times {\mathbf{ d}}^{*}=0$, which denotes that the gap matrix is unitary \cite{superconductivity91}.

The wave function in the NM region can be expressed as \cite{NMwavefunction15}
\begin{equation}
    \Phi_{\text{N}\alpha}({\mathbf{r}})=\sum_{|{{\mathbf{k}}_{\parallel}}|<k_{F}}\left[
\begin{pmatrix}
\delta_{\alpha\uparrow} \\
\delta_{\alpha\downarrow} \\
0 \\
0
\end{pmatrix}e^{ik_zz}+
\begin{pmatrix}
b_{\alpha\uparrow} \\
b_{\alpha\downarrow} \\
0 \\
0
\end{pmatrix}e^{-ik_zz}+
\begin{pmatrix}
0 \\
0 \\
a_{\alpha\uparrow} \\
a_{\alpha\downarrow}
\end{pmatrix}e^{ik_zz}\right] \cdot f_{{\mathbf{k}}_{\parallel}}({\boldsymbol{\rho}}),\label{wavefunction in NM}
\end{equation}
where $\delta_{\alpha\beta}$ is the Kronecker symbol, while $a_{\alpha\beta}$ and $b_{\alpha\beta}$ stand for the Andreev reflection and normal reflection coefficients, respectively [Supplementary Fig.~5(b)]. The subscript indices $\alpha$ and $\beta$ denote the spin state. The spatial wave function parallel to the junction interface is expressed as $f_{{\mathbf{k}}_{\parallel}}({\boldsymbol{\rho}})=e^{i {\mathbf{ k}}_{\parallel}\cdot{\boldsymbol{\rho}}}/\sqrt{S}$, where $S$ is area of the interface. The displacement vector ${\boldsymbol{\rho}}=(x,~y,~0)$ lies within the interfacial plane and the wave vector component parallel to the junction interface is given by ${\mathbf{k}}_{\parallel}=(k_x,~k_y,~0)$. For those particles with energy $E\ll\varepsilon_F$, the wave vector components obey the approximate relation $k_z^2+|{\mathbf{k}}_{\parallel}|^2\approx k_F^2=\sqrt{\frac{2m\varepsilon_F}{\hbar^2}}$.

The wave function in the SC is given by \cite{superconductivity91}
\begin{align}
    \Phi_{\text{S}\alpha}({\mathbf{r}})&=\sum_{|{\mathbf{k}}_{\parallel}|<k_{F}}\left[c_{\alpha1}^{(e)}
\begin{pmatrix}
u_{+} \\
0 \\
-\tilde{d}_{1,+}^{*}-i\tilde{d}_{2,+}^{*} \\
\tilde{d}_{3,+}^{*}
\end{pmatrix}e^{ik_zz}+c_{\alpha2}^{(e)}
\begin{pmatrix}
0 \\
u_{+} \\
\tilde{d}_{3,+}^{*} \\
\tilde{d}_{1,+}^{*}-i\tilde{d}_{2,+}^{*}
\end{pmatrix}e^{ik_zz}\right.\nonumber\\
&\left.+c_{\alpha1}^{(h)}
\begin{pmatrix}
-\tilde{d}_{1,-}+i\tilde{d}_{2,-} \\
\tilde{d}_{3,-} \\
u_{-} \\
0
\end{pmatrix}e^{-ik_zz}+c_{\alpha2}^{(h)}
\begin{pmatrix}
\tilde{d}_{3,-} \\
\tilde{d}_{1,-}+i\tilde{d}_{2,-} \\
0 \\
u_{-}
\end{pmatrix}e^{-ik_zz}\right] \cdot f_{{\mathbf{k}}_{\parallel}}({\boldsymbol{\rho}}),\label{wavefunction in SC}
\end{align}
where
\begin{align}
   &u_{\pm}=\frac{E+\xi_{\pm}}{\sqrt{(E+\xi_{\pm})^{2}+|{\mathbf{ d}}_{\pm}|^2}},~\tilde{d}_{i,\pm}=\frac{d_{i,\pm}}{\sqrt{(E+\xi_{\pm})^2+|{\mathbf{d}}_{\pm}|^2}}, \nonumber\\
   &\xi_{\pm}=\sqrt{E^2-|{\mathbf{d}}_{\pm}|^2},~{\mathbf{d}}_{\pm}={\mathbf{d}}({\pm k_z},{\mathbf{k}}_{\parallel})=(d_{1,\pm},~d_{2,\pm},~d_{3,\pm}), \nonumber
\end{align}
and $c_{\alpha i}^{(e(h))}$ represents the transmission coefficient for an electron-like (hole-like) quasiparticle. Notably, the kinetic energy of quasiparticle, $\xi=\sqrt{E^2-|{\mathbf{d}}({\mathbf{ k})}|^2}$, satisfies the following relation, due to the causality of the transport coefficients \cite{causality21}
\begin{align}
    \xi=
\begin{cases}
\sqrt{E^2-|{\mathbf{d}}({\mathbf{k})}|^2},~~~~(E\geq|{\mathbf{d}}({\mathbf{k})}|)\\
i\sqrt{|{\mathbf{d}}({\mathbf{k})}|^2-E^2},~~~~(-|{\mathbf{d}}({\mathbf{k})}|<E<|{\mathbf{d}}({\mathbf{k})}|) \\
-\sqrt{E^2-|{\mathbf{d}}({\mathbf{k})}|^2}.~~~~(E\leq-|{\mathbf{d}}({\mathbf{k})}|) &
\end{cases}
\end{align}

The boundary conditions for the wave functions are \cite{causality21}
\begin{align}
    \Phi_{\text{S}\alpha}({\mathbf{r}})|_{z=0^+}=\Phi_{\text{N}\alpha}({\mathbf{r}})|_{z=0^-}&,\label{b1}\\
    -\frac{\hbar^2}{2m}\left[\partial_z\Phi_{\text{S}\alpha}({\mathbf{r}})|_{z=0^+}-\partial_z\Phi_{\text{N}\alpha}({\mathbf{r}})|_{z=0^-}\right]+V_0&\Phi_{\text{S}\alpha}({\mathbf{r}})|_{z=0^+}=0,\label{b2}
\end{align}
where $m$ is the mass of the electron. The boundary conditions are derived from the continuity of the wave function and the conservation of the probability current density, respectively.

Substituting the wave functions Eqs. (\ref{wavefunction in NM}) and (\ref{wavefunction in SC}) into two boundary conditions, we can derive the following equations:
\begin{align}
    \begin{pmatrix}
        \hat{\sigma}_0\\
        \hat{R}_{h}
    \end{pmatrix}=\check{U}
    \begin{pmatrix}
        (1+i\frac{Z}{\tilde{k}})\hat{T}_{e}\\
        i\frac{Z}{\tilde{k}}\hat{T}_{h}
    \end{pmatrix},\label{R_h eq}\\
    \begin{pmatrix}
        \hat{R}_{e}\\
        0
    \end{pmatrix}=\check{U}
    \begin{pmatrix}
        -i\frac{Z}{\tilde{k}}\hat{T}_{e}\\
        (1-i\frac{Z}{\tilde{k}})\hat{T}_{h}
    \end{pmatrix},\label{R_e eq}
\end{align}
where \begin{align}
    &\check{U}=\begin{bmatrix}
        u_{+}\hat{\sigma}_0 & \hat{v}_{-}\\
        \hat{v}_{+} & u_{-}\hat{\sigma}_0
    \end{bmatrix},~\hat{v}_{+}=\frac{\hat{\Delta}^{\dagger}_{+}}{\sqrt{(E+\xi_{+})^2+|{\mathbf{d}}_{+}|^2}},\nonumber\\
    &\hat{v}_{-}=\frac{\hat{\Delta}_{-}}{\sqrt{(E+\xi_{-})^2+|{\mathbf{d}}_{-}|^2}},~\hat{\Delta}_{\pm}=\hat{\Delta}(\pm k_z,{\mathbf{k}}_{\parallel}),~Z=\frac{mV_0}{\hbar^2k_F},~\tilde{k}=\frac{k_z}{k_F}.\nonumber
\end{align}
$\hat{R}_{h}=\begin{bmatrix}
    a_{\uparrow\uparrow} & a_{\downarrow\uparrow}\\
    a_{\uparrow\downarrow} & a_{\downarrow\downarrow}
\end{bmatrix}$ is the Andreev reflection matrix, $\hat{R}_{e}=\begin{bmatrix}
    b_{\uparrow\uparrow} & b_{\downarrow\uparrow}\\
    b_{\uparrow\downarrow} & b_{\downarrow\downarrow}
\end{bmatrix}$ is the normal reflection matrix, and $\hat{T}_{e(h)}=\begin{bmatrix}
    c_{\uparrow1}^{(e(h))} & c_{\downarrow1}^{(e(h))}\\
    c_{\uparrow2}^{(e(h))} & c_{\downarrow2}^{(e(h))}
\end{bmatrix}$ is the transmission matrix of electron-like (hole-like) quasiparticles, respectively.

By solving Eqs.~\eqref{R_h eq} and \eqref{R_e eq}, we can get solutions for the reflection matrix:
\begin{equation}
        \hat{R}_{e}= r_{n}\left(1-\frac{\hat{v}_{-}\hat{v}_{+}}{u_{+}u_{-}}\right)\hat{A}^{-1},~\hat{R}_{h}=|t_{n}|^{2}\frac{\hat{v}_{+}}{u_{+}}\hat{A}^{-1}
\end{equation}
where $r_n=\frac{-iZ}{\tilde{k}+iZ},~t_n=\frac{\tilde{k}}{\tilde{k}+iZ},~\hat{A}=1-|r_{n}|^2\frac{\hat{v}_{-}\hat{v}_{+}}{u_{+}u_{-}}$.

Total current across the NM-SC junction deduced from the Landauer-B\"{u}ttiker expression is given by \cite{transport95}
\begin{equation}
    I_\text{NS}=I_0\int_{-\infty}^{+\infty}\left[f_{L}(E)-f_{R}(E)\right]\sum_{{\mathbf{k}}_{\parallel}}\left\{1+\frac{1}{2}\sum_{\alpha,\beta}\left[|a_{\alpha\beta}(E)|^2-|b_{\alpha\beta}(E)|^2\right]\right\}{\rm d}E,
\end{equation}
where $f_{L(R)}(E)=\left(1+e^{E-\mu_{L(R)}/k_BT}\right)^{-1}$ is the Fermi-Dirac distribution function of the left (right) reservoirs with the chemical potentials $\mu_{L}=\varepsilon_F+eV$ and $\mu_R=\varepsilon_F$, respectively.

The reflection coefficients satisfy
\begin{equation}
    \sum_{\alpha,\beta}\{|b_{\alpha\beta}(E)|^2={\rm Tr}(\hat{R}_{e}\hat{R}_{e}^{\dagger}),~\sum_{\alpha,\beta}\{|a_{\alpha\beta}(E)|^2={\rm Tr}(\hat{R}_{h}\hat{R}_{h}^{\dagger}).
\end{equation}

Then, with the Andreev reflection and normal reflection coefficients, the differential conductance (as a function of $V$ and $T$) is given by
\begin{equation}
    G_\text{NS}(V,T)=\frac{{\rm d}}{{\rm d}V}I_\text{NS}=G_0\int_{-\infty}^{+\infty}\frac{\partial f(E-eV)}{\partial V}\sum_{{\mathbf{k}}_{\parallel}}\left\{1+\frac{1}{2}\sum_{\alpha,\beta}\left[|a_{\alpha\beta}(E)|^2-|b_{\alpha\beta}(E)|^2\right]\right\}{\rm d}E,
\end{equation}

Specially, at zero temperature, the differential conductance is obtained,
\begin{equation}
    G_\text{NS}(V)|_{T=0}=G_0\left\{1+\frac{1}{2}\sum_{\alpha,\beta}\left[|a_{\alpha\beta}(eV)|^2-|b_{\alpha\beta}(eV)|^2\right]\right\}.
\end{equation}

A detailed solution of the BdG equations in the presence of inelastic scattering \cite{Gamma94}
shows that the impact of finite quasiparticle lifetime can be effectively modeled through
the substitution $E \rightarrow E+i\Gamma$. This leads us to the final expression for the differential conductance in this model,
\begin{align}
    G_\text{NS}(V,T)=G_0\int_{-\infty}^{+\infty}\frac{\partial f(E-eV)}{\partial V}\sum_{{\mathbf{k}}_{\parallel}}\left\{1+\frac{1}{2}\sum_{\alpha,\beta}\left[|a_{\alpha\beta}(E+i\Gamma)|^2-|b_{\alpha\beta}(E+i\Gamma)|^2\right]\right\}{\rm d}E.
\end{align}


\newpage

\textbf{SI \Rmnum{5}: Simulations of Andreev reflection PCS for spin-triplet superconductivity}\\

Supplementary Figure 6 shows the simulated PCS at different temperatures, based on $p_{x(y)}$- (left column) and $p_z$-(right column) wave pairings. For comparison, all the simulations were performed with fixed $\Gamma/\Delta=0.2$. For small $Z=0.2$, both $p_{x(y)}$ and $p_z$ waves show peak-like feature at zero bias, as shown in Supplementary Fig.~6(a,c). For large $Z=1$, they differ to each other substantially. For $p_{x(y)}$ wave, a zero-bias dip develops [Supplementary Fig.~6(a)]. In contrast, for $p_{z}$ wave, the ZBCP is robust against varying $Z$, but meanwhile a dip develops on each side of the ZBCP, cf. Supplementary Fig.~6(d). Another major difference between $p_{x(y)}$ and $p_{z}$ waves lies in the amplitude of the ZBCP, with the former obviously smaller than the latter, as is mentioned in the main text.

\begin{figure*}[!htp]
	\vspace*{-0pt}
	\includegraphics[width=15cm]{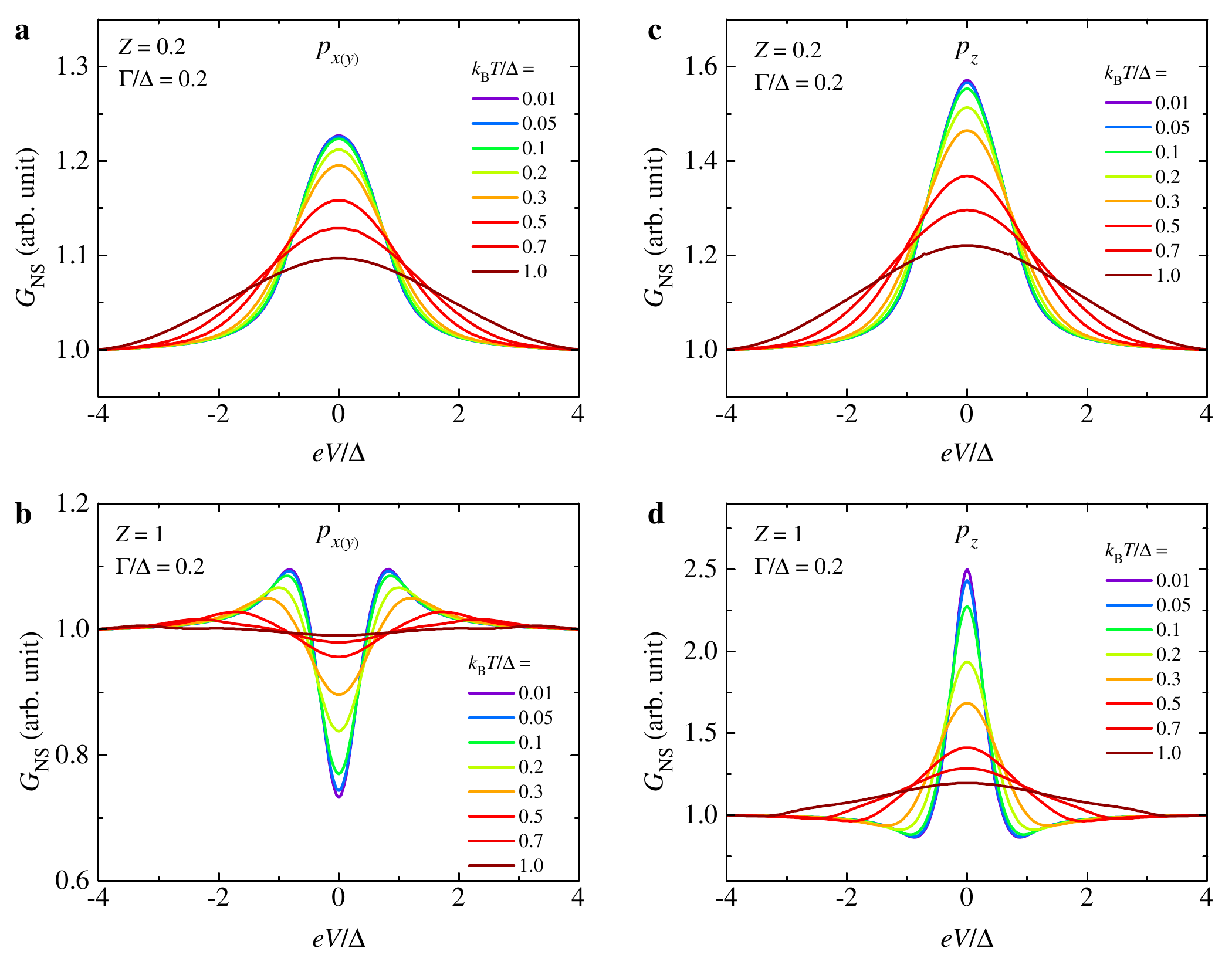}
	\label{Fig.S6}
\vspace*{-0pt}
\small
\begin{flushleft}
\justifying{
    \textbf{Supplementary Figure 6 $|$ Simulated PCS at different temperatures.} The NM-SC junction is on the $\mathbf{x}$-$\mathbf{y}$ facet. \textbf{a} $p_{x(y)}$, $Z=0.2$, $\Gamma/\Delta=0.2$. \textbf{b} $p_{x(y)}$, $Z=1$, $\Gamma/\Delta=0.2$. \textbf{c} $p_z$, $Z=0.2$, $\Gamma/\Delta=0.2$. \textbf{d} $p_z$, $Z=1$, $\Gamma/\Delta=0.2$. It is found that the behavior of ZBCP with a side-dip is present in $p_z$-wave pairing but absent in $p_{x(y)}$-wave pairing symmetry.
    }
\end{flushleft}
\normalsize
\end{figure*}

\begin{figure*}[!htp]
	\vspace*{0pt}
	\includegraphics[width=15cm]{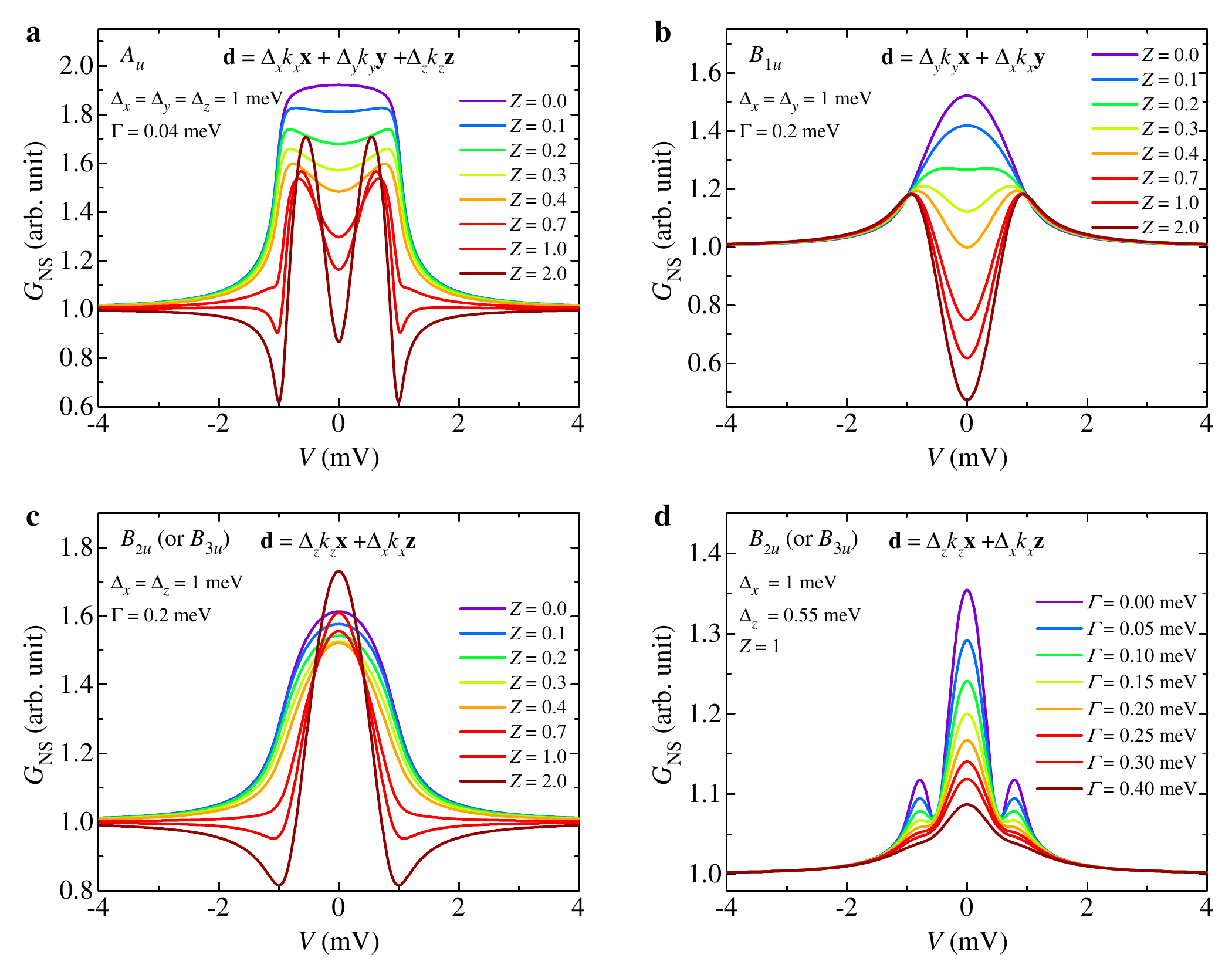}
	\label{Fig.S7}
\vspace*{-0pt}
\small
\begin{flushleft}
\justifying{
    \textbf{Supplementary Figure 7 $|$ Simulated PCS for various irreducible representations.} \textbf{a} $A_u$. \textbf{b} $B_{1u}$. \textbf{c}-\textbf{d} $B_{2u}$ or $B_{3u}$. The simulations are made at $T=0.3$ K. Note that for the junction within the $\mathbf{x}$-$\mathbf{y}$ plane, the results of $B_{2u}$ and $B_{3u}$ are the same. For $A_u$, a zero-bias dip appears for all $Z>0$, and a side dip appears for large $Z$. For $B_{1u}$, a zero-bias dip appears for large $Z$, but no side dip can be obtained. For $B_{2u}$ and $B_{3u}$, the ZBCP persists for all $Z$, and a dip on each side appears for large $Z$. As $\Gamma$ decreases, the shoulder may evolves into a satellite peak.
    }
\end{flushleft}
\normalsize
\end{figure*}

We then simulate the PCS for different representations, viz $A_u$, $B_{1u}$, $B_{2u}$ and $B_{3u}$, and the results are displayed in Supplementary Fig.~7. The basis functions of these representations have been listed in Table 1. Note that here we have ignored the high-order ($k_xk_yk_z$) terms whose effect on PCS turns out to be negligible, cf Supplementary Fig.~13. The main parameters used are listed in the figures. For $A_u$ presentation, a ``flattened" ZBCP shows up for $Z=0$, but once $Z>0$, the peak gradually splits into two peaks; a dip appears at $|V|>0$ when $Z>1$, as can be seen in Supplementary Fig.~7(a). The sensitivity of the ZBCP to the values of $Z$ and $\Gamma$ is depicted by the contour plot of $\mathrm{d}^2G_\text{NS}(0)/\mathrm{d}V^2$ as shown in Supplementary Fig.~8. For $B_{1u}$ presentation, a ZBCP is also visible for small $Z$, but as $Z$ increases, it gradually evolves into a zero-bias dip [Supplementary Fig.~7(b)]. As is shown in Supplementary Fig.~7(c,d), the ZBCP in the $B_{2u}$ presentation is robust to the variations of $Z$ and $\Gamma$. In contrast, the finite-bias conductance is highly susceptible. For fixed $\Gamma=0.05$ meV, side dips emerge as $Z$ exceeds 0.8. Furthermore, at $Z=1$, a large $\Gamma$ of 0.40 meV produces only a shallow, shoulder-like dip. This dip deepens and sharpens into well-defined satellite peaks as $\Gamma$ is reduced. The $B_{3u}$ behaves similarly to $B_{2u}$. We shall see that $B_{2u}$ and $B_{3u}$ capture all the main features of the experimental PCS of UTe$_2$.

\begin{figure*}[!htp]
	\vspace*{-0pt}
	\includegraphics[width=14cm]{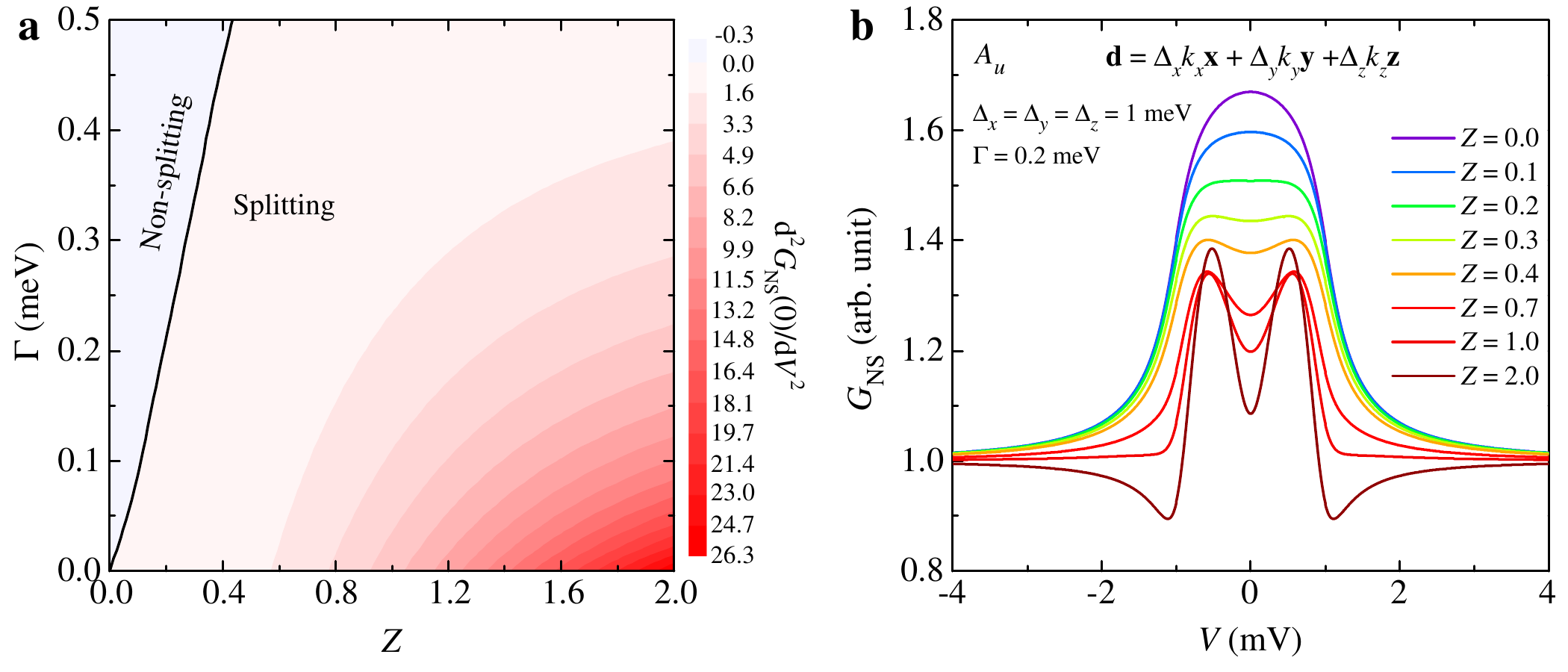}
	\label{Fig.S8}
\vspace*{-10pt}
\small
\begin{flushleft}
\justifying{
    \textbf{Supplementary Figure 8 $|$ Additional simulations for $A_u$.} The parameters used are: $\Delta_x = \Delta_y = \Delta_z = 1$ meV, and $T = 0.3$ K. \textbf{a} Contour plot of $\mathrm{d}^2G_\text{NS}(0)/\mathrm{d}V^2$ as a function of $Z$ and $\Gamma$. The black line is the boundary of the regions where $G_\text{NS}(V)$ does and doesn¨t split at zero bias. For any reasonably large $\Gamma$, $G_\text{NS}(V)$ either splits or shows a flat and broad central peak at $V=0$, as exemplified by the simulations with fixed $\Gamma = 0.2$ meV and variable $Z$ in \textbf{b}. These features are qualitatively inconsistent with our experiments.
    }
\end{flushleft}
\normalsize
\end{figure*}

\clearpage

\textbf{SI \Rmnum{6}: Fitting of the experimental data}\\

\begin{table*}[!htp]
\begin{flushleft}
\justifying{
    Supplementary Table 1 : Fitting parameters of PCS at 0.3 K to $B_{2u}$ (or $B_{3u}$) representation, for samples S1-S3 at $p=0$ and $B=0$.
    }
\end{flushleft}
\begin{ruledtabular}
\begin{tabular}{ccccc}
 Sample   & $\Delta_{x(y)}$ (meV)  &  $\Delta_{z}$ (meV)  &  $Z$      &   $\Gamma$ (meV)   \\ \hline
   S1     & 1.18(7)             &   0.72(4)               &  1.67(9)  &    0.39(2)         \\
   S2     & 1.15(1)             &   0.62(1)               &  0.76(1)  &    0.40(1)         \\
   S3     & 1.15(1)             &   0.60(1)               &  1.12(1)  &    0.16(1)         \\
\end{tabular}
\end{ruledtabular}
\small
\vspace*{-10pt}
\begin{flushleft}
\end{flushleft}
\normalsize
\end{table*}

\begin{figure*}[!htp]
	\vspace*{-0pt}
	\includegraphics[width=16.5cm]{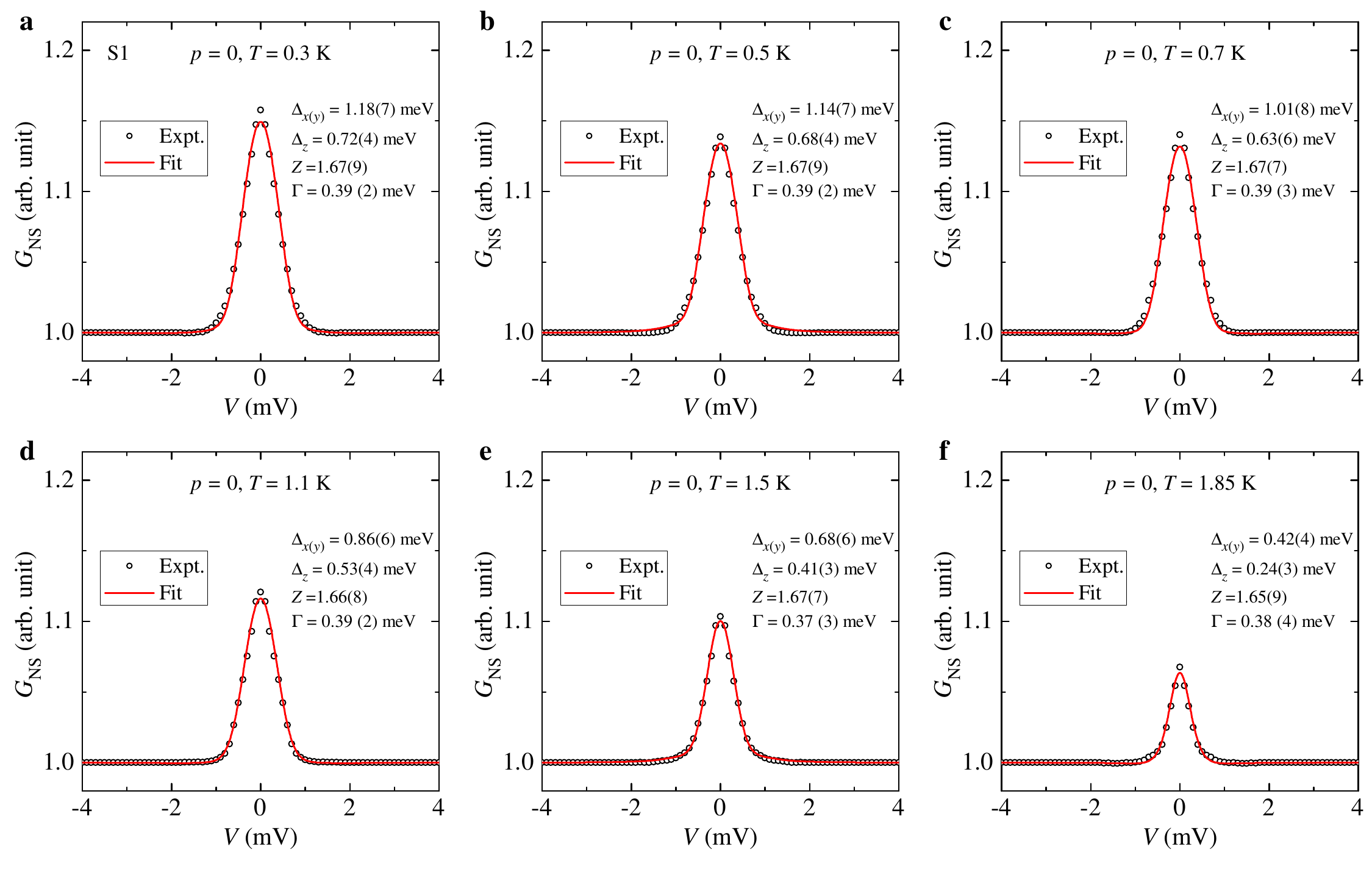}
	\label{Fig.S9}
\vspace*{-20pt}
\small
\begin{flushleft}
\justifying{
    \textbf{Supplementary Figure 9 $|$ Fitting of the ambient-pressure PCS for sample S1.} \textbf{a} 0.3 K. \textbf{b} 0.5 K. \textbf{c} 0.7 K. \textbf{d} 1.1 K. \textbf{e} 1.5 K. \textbf{f} 1.85 K. The fittings were carried out with $B_{2u}$ (or $B_{3u}$). The fitting parameters are listed in each panels.
    }
\end{flushleft}
\normalsize
\end{figure*}

\begin{figure*}[!htp]
	\vspace*{-0pt}
	\includegraphics[width=16.5cm]{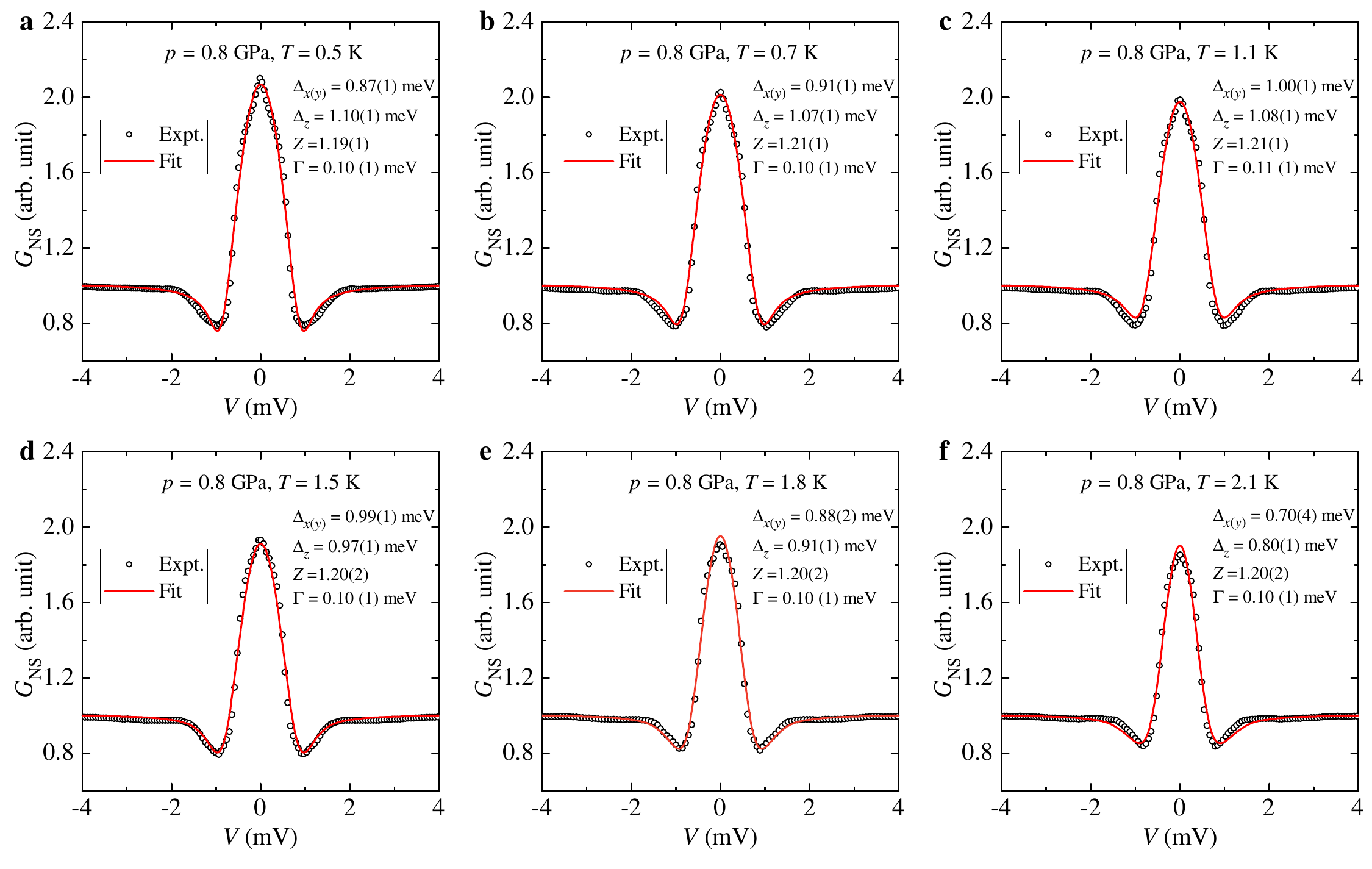}
	\label{Fig.S10}
\vspace*{-25pt}
\small
\begin{flushleft}
\justifying{
    \textbf{Supplementary Figure 10 $|$ Fitting of the point-contact spectra at 0.8 GPa for sample S1.} \textbf{a} 0.5 K. \textbf{b} 0.7 K. \textbf{c} 1.1 K. \textbf{d} 1.5 K. \textbf{e} 1.8 K. \textbf{f} 2.1 K. The fittings were carried out with $B_{2u}$ (or $B_{3u}$).
    }
\end{flushleft}
\normalsize
\end{figure*}

\begin{figure*}[!htp]
	\vspace*{-0pt}
	\includegraphics[width=10cm]{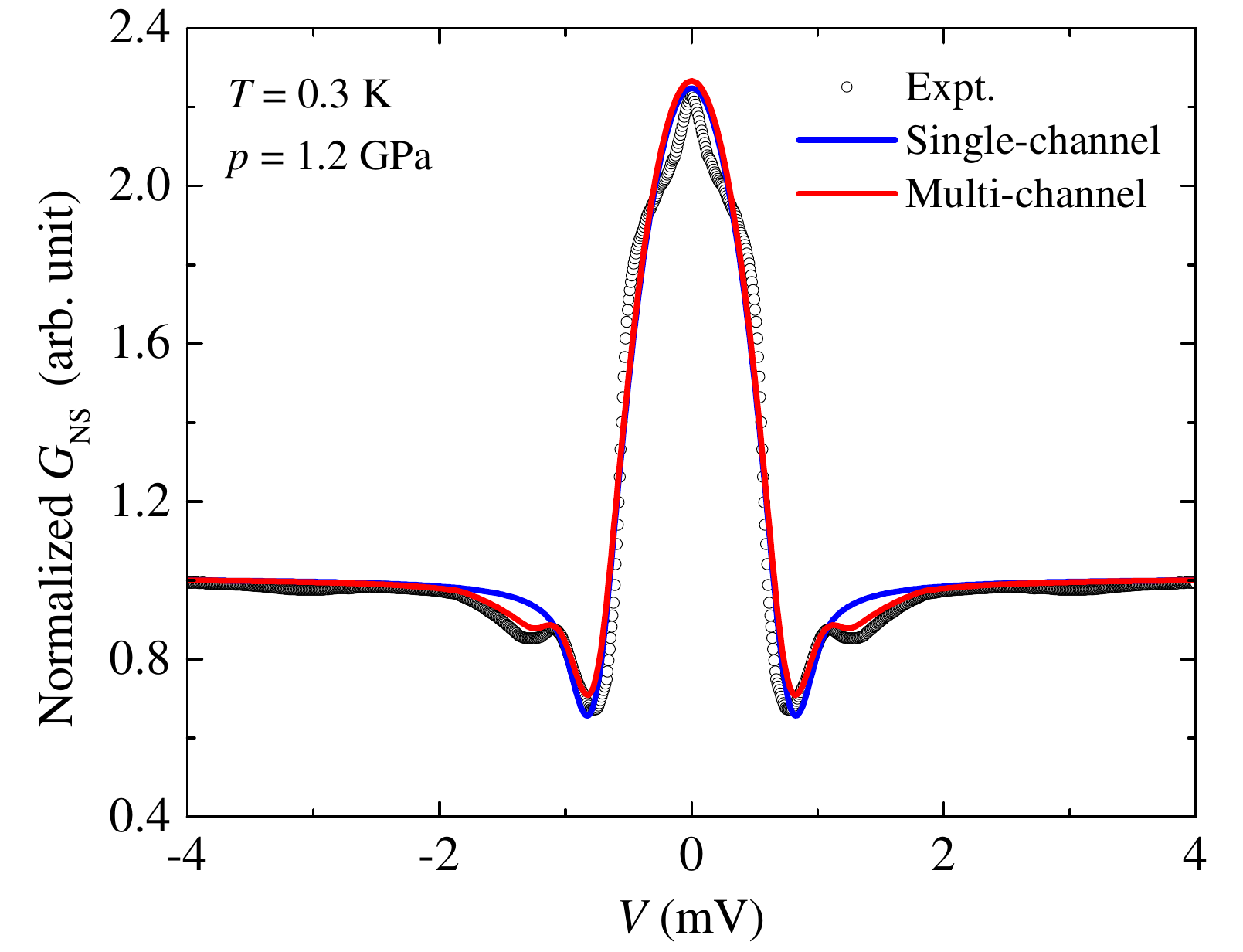}
	\label{Fig.S11}
\vspace*{-5pt}
\small
\begin{flushleft}
\justifying{
    \textbf{Supplementary Figure 11 $|$ Fitting of the point-contact spectra at 1.2 GPa for sample S1.} Taking the data of 0.3 K as an example. It is found that the single-channel BTK model can not capture the double-side-dip feature at this pressure.
    }
\end{flushleft}
\normalsize
\end{figure*}

\begin{figure*}[!htp]
	\vspace*{-0pt}
	\includegraphics[width=16.5cm]{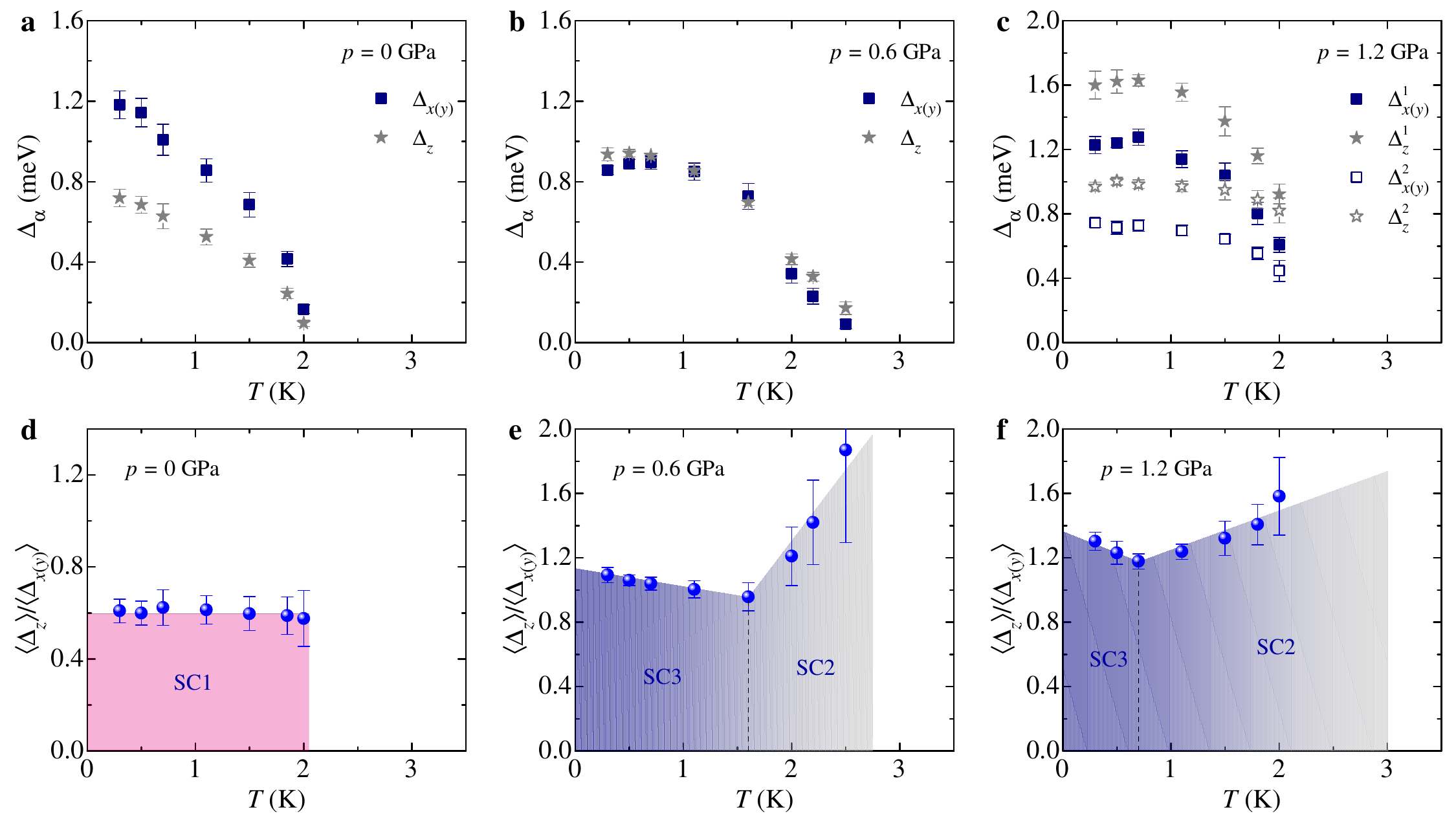}
	\label{Fig.S12}
\vspace*{-20pt}
\small
\begin{flushleft}
\justifying{
    \textbf{Supplementary Figure 12 $|$ The obtained superconducting gap parameters for sample S1.} \textbf{a}-\textbf{c} $\Delta_\alpha$ [$\alpha=x(y), z$] as a function of temperature for $p=$ 0, 0.6 and 1.2 GPa. \textbf{d}-\textbf{f} $\langle \Delta_{z}\rangle/\langle\Delta_{x(y)}\rangle$.
    }
\end{flushleft}
\normalsize
\end{figure*}

\newpage
\textbf{SI \Rmnum{7}: Consideration of high-order term in SC order parameter}\\

Symmetry allows higher-order basis functions -- e.g., the $k_x k_y k_z$ term -- in the $B_{1u}$ pairing gap, whose inclusion will also cause a sign change in $\Delta(\mathbf{k})$ upon going from $k_z$ to $-k_z$. To estimate how this term affects the PCS for tunneling current $\mathbf{I}\parallel\mathbf{c}$, we carried out additional calculations for the $B_{1u}$ state. Assuming the high-order component has a large amplitude comparable to the linear-order ones, we set $\mathbf{d}=\Delta (k_y,~k_x,~k_x k_y k_z)$ and compared with the simplified $B_{1u}$ form $\mathbf{d} = \Delta (k_y,~k_x,~0)$. $k_F=1$ is used in the simulations. As shown in Supplementary Fig.~13, the inclusion of the $k_x k_y k_z$ term produces only minor modifications to the calculated spectra. In particular, this extended $B_{1u}$ model still cannot account for the prominent side-dip/three-peak features seen in the experiments.

\begin{figure*}[!htp]
	\vspace*{-10pt}
	\includegraphics[width=14cm]{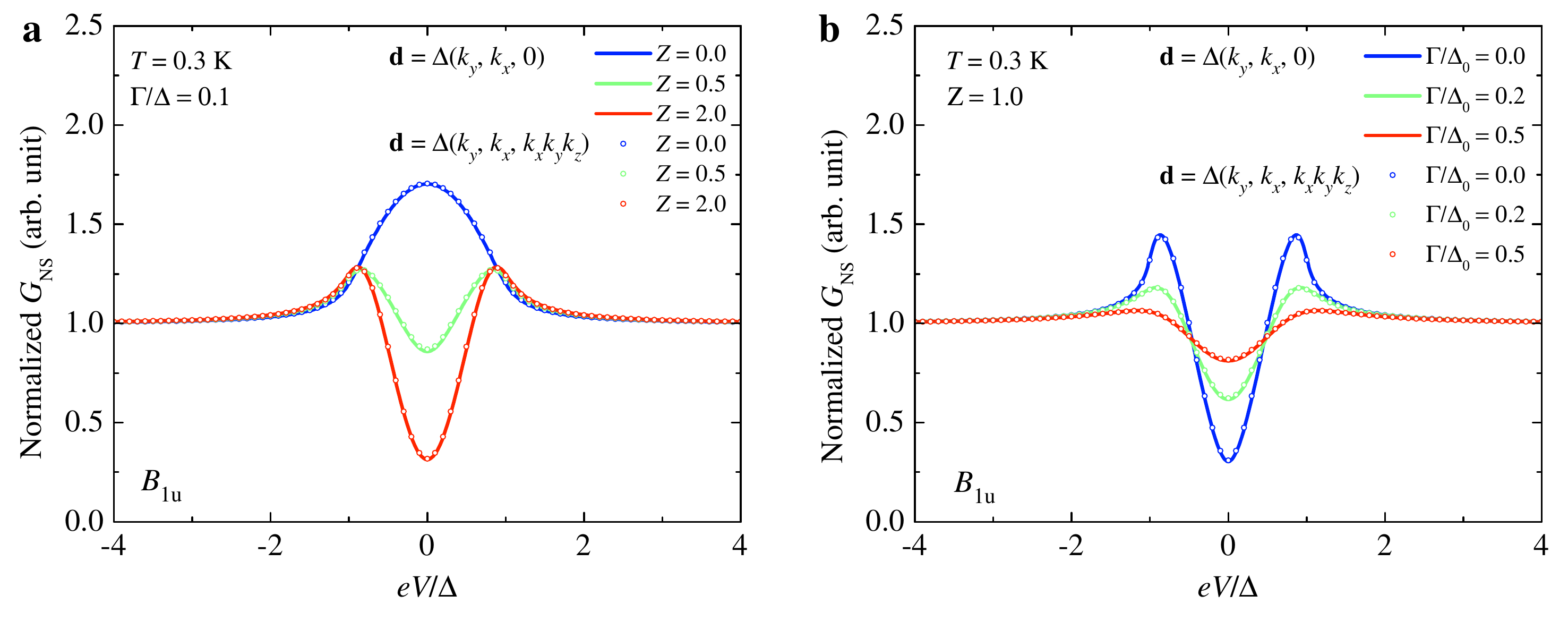}
	\label{Fig.S13}
\vspace*{-10pt}
\small
\begin{flushleft}
\justifying{
    \textbf{Supplementary Figure 13 $|$ Comparison of point-contact spectra for the $B_{1u}$ pairing with and without the high-order $k_x k_y k_z$ term.} $k_F=1$ in these calculations. For simplified $B_{1u}$, $\mathbf{d}=\Delta(k_y,~k_x,~0)$, and the results are represented by open circles; for high-order imposed $B_{1u}$, $\mathbf{d}=\Delta(k_y,~k_x,~k_x k_y k_z)$, and the results are depicted by solid lines. \textbf{a} Simulations with fixed $\Gamma/\Delta=0.1$, and varying $Z$ = 0, 0.5 and 2.0. \textbf{b} Simulations with fixed $Z=1.0$, and varying $\Gamma/\Delta$ = 0, 0.2 and 0.5.
    }
\end{flushleft}
\normalsize
\end{figure*}

\newpage
\textbf{SI \Rmnum{8}: Consideration of anisotropic Fermi surface}\\

In this section, we performed some simulations aiming at comparing the conductance spectra for isotropic and anisotropic Fermi surfaces, seeing Supplementary Fig.~14. The extent of FS anisotropy is controlled by the parameter $r_v\equiv v_F^c/v_F^{ab}$, where $v_F^{ab}$ and $v_F^{c}$ are the in-plane and out-of-plane Fermi velocities, respectively. While the lineshape of $G_\text{NS}(V)$ changes with $r_v$, the behavior
near the gap edge remains essentially unchanged.

\begin{figure*}[!htp]
	\vspace*{-0pt}
	\includegraphics[width=14cm]{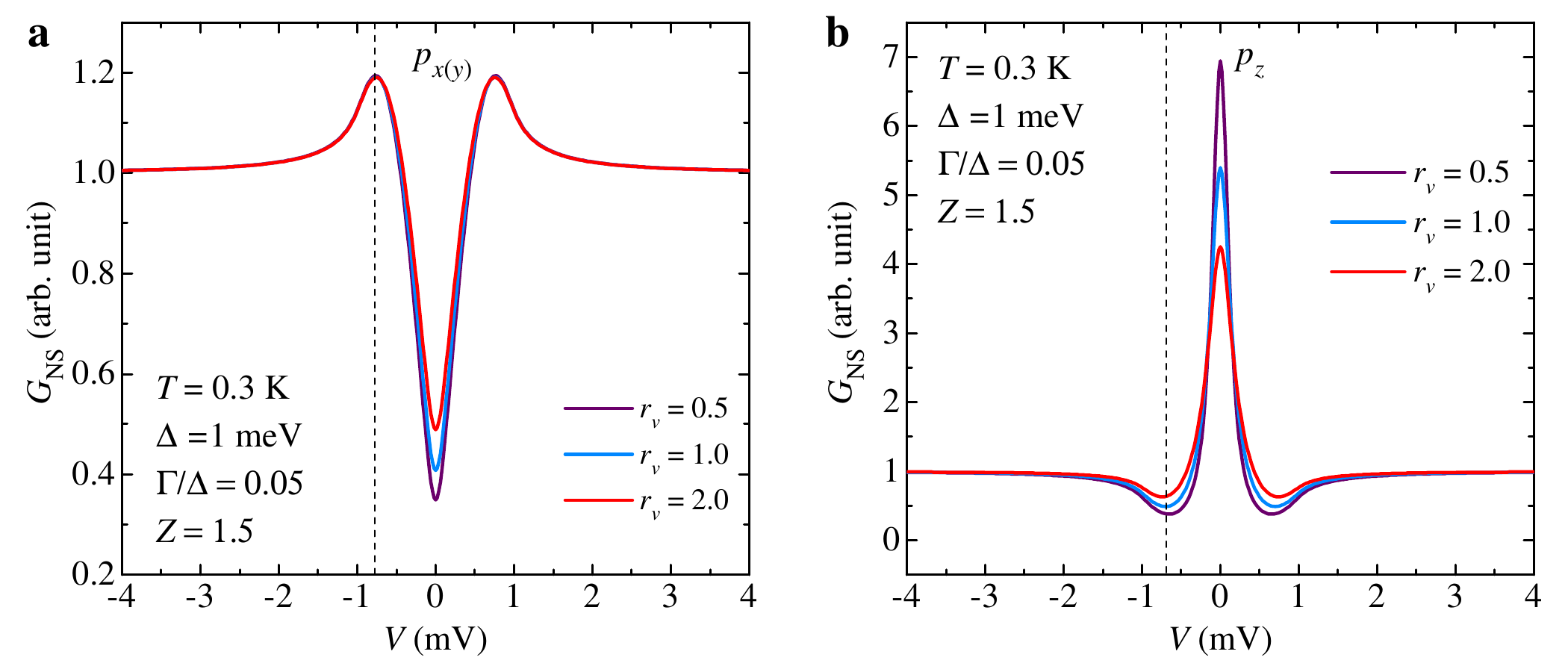}
	\label{Fig.S14}
\vspace*{-10pt}
\small
\begin{flushleft}
\justifying{
    \textbf{Supplementary Figure 14 $|$ Simulations of point-contact spectra with varying Fermi surface anisotropy.} The ratio $r_v\equiv v_F^c/v_F^{ab}$, with $v_F^{ab}$ and $v_F^{c}$ being the in-plane and out-of-plane Fermi velocities, respectively.
    }
\end{flushleft}
\normalsize
\end{figure*}

\end{document}